\documentstyle[12pt]{article}
\newcommand{\e}{\; {\rm e} }
\newcommand{\sgn}{\; {\rm sgn} }
\newcommand{\tr}{\; {\bf\rm tr} }
\newcommand{\ch}{\; {\rm ch} }
\newcommand{\th}{\; {\rm th} }
\newcommand{\sh}{\; {\rm sh} }

\newcommand{\be}{\begin{eqnarray} }
\newcommand{\ee}{\end{eqnarray} }

\begin{document}

\begin{center}
\Large{\bf Topological Effects in Medium\\}
\vskip 5mm
\large{ A. N. Sissakian\\}
{\it Bogolubov Theoretical Laboratory,
Joint Institute for Nuclear Research,
Dubna, Moscow region 141980, Russia\\}
\vskip 5mm
\large{O. Yu. Shevchenko\footnote{shevch@nusun.jinr.ru}
and S. B. Solganik\footnote{solganik@thsun1.jinr.ru}\\}
{\it Laboratory of Nuclear Problems,
Joint Institute for Nuclear Research,
Dubna, Moscow region 141980, Russia\\}
\end{center}
\vskip 5mm

\begin{abstract}
Two closely related topological phenomena are studied
at finite density and temperature.
These are chiral anomaly and Chern--Simons term.
It occurs that the chiral anomaly doesn't depend on
density and temperature.
Chern-Simons term appearance in even dimensions is studied
under two types of constraints: chiral and usual charges conservation.
In odd dimensions, by using different methods
it is shown  that $\mu^2 = m^2$ is the
crucial point for Chern--Simons at zero temperature.
So when $\mu^2 < m^2$  $\mu$--influence disappears and
we get the usual Chern-Simons term.
On the other hand, when $\mu^2 > m^2$ the Chern-Simons term
vanishes because of non--zero density of background fermions.
The connection between parity anomalous Chern-Simons in odd dimension and
chiral anomaly in even dimension is established at arbitrary density
and temperature.
These results hold in any  dimension as in abelian,
so as in nonabelian cases.
\\
\end{abstract}
\vskip 5mm

\section{Introduction.}
There is a lot of physical processes where density and temperature
play essential role. These are processes occurred under large
density background, for example, in quark--gluon plasma or in
neutron stars. On the other hand, there exist processes where
even negligible density or temperature may give rise to principal
effects.
One of the most interesting areas, where density and temperature
influence could be considerable, is the area of topological effects.
Here, even
negligible density or temperature could change the topology of the
a problem as whole, what could lead to  considerable influence.
In particular, here we are interested in Chern-Pontriagin and
Chern-Simons  secondary characteristic classes.
That corresponds to chiral anomaly in even dimensions and
to Chern-Simons (parity anomaly) in odd dimensions.
Both phenomena are very important in quantum physics.
So, chiral anomalies in quantum field theory have certain direct
applications to the decay of $\pi_{0}$ into two photons
($\pi_{0}\rightarrow\gamma\gamma$), in the understanding and solution
of the U(1) problem and so on.
On the other hand, there are many
effects caused by Chern-Simons secondary
characteristic class. These
are, for example, gauge particles mass appearance  in quantum
field theory, applications to condense matter physics  such as
the fractional quantum Hall effect and high $T_{c}$ superconductivity,
possibility of free of metric tensor theory construction etc.

It must be emphasized that these two phenomenons are closely
related. As it was shown (at zero density) in \cite{ni,ni2,ni1}
the trace identities
connect even dimensional anomaly with the odd dimensional
Chern-Simons.
The main goal of this article is to consider these  anomalous
objects  at finite density and temperature.

It was shown \cite{niemi,redl,witten} in a conventional zero density
and temperature gauge
theory that the Chern-Simons term is generated in the
Eulier--Heisenberg effective action
by quantum corrections.
Since  the chemical potential term $\mu\bar\psi\gamma^{0}\psi$ is
odd under charge conjugation we can expect that it would
contribute to $P$ and $CP$ nonconserving quantity ---  Chern-Simons term.
As we will see, this expectation is completely justified.
The zero density approach usually is a good quantum field  approximation
when the chemical potential is small as compared with
characteristic energy scale  of physical processes.
Nevertheless, for investigation of topological effects
it is not the case.
As we will see below, even a small density could lead to
principal effects.

In the excellent paper by Niemi \cite{ni} it was emphasized that the
charge density
at $\mu \not = 0$ becomes nontopological object, i.e contains as topological
part so as nontopological one.
The charge density at $\mu \not = 0$ (nontopological, neither parity odd
nor parity even object)\footnote{For abbreviation,  speaking about parity
invariance properties of local objects, we will
keep in mind  symmetries of the corresponding action parts.}
in $QED_{3}$ at finite density
was calculated and exploited in \cite{tolpa}. It must be
emphasized that in \cite{tolpa} charge density
(calculated in the constant pure magnetic field)
contains
as well parity odd part corresponding to Chern-Simons term,
so as parity even part, which can't be covariantized
and don't contribute to the mass of the gauge field.
Here we are interested in  finite density and temperature
influence
on covariant parity odd form in action leading to the
gauge field mass generation --- Chern-Simons
topological term. Deep insight on this phenomena at small densities
was done in \cite{ni,ni1}.
The result for Chern-Simons term coefficient in $QED_{3}$  is
$$\left[ \th \frac{1}{2}\beta(m-\mu)+\th \frac{1}{2}\beta(m+\mu)\right],$$
see \cite{ni1}, formulas (10.18).
However, to get this result it was heuristicaly supposed
that at small densities index theorem could still be used and
only odd in energy part of spectral density is responsible for
parity nonconserving effect.
Because of this in \cite{ni1} it had been stressed
that the result holds only for small $\mu$. However,
as we'll see below this result holds for any values of
chemical potential.
Thus, to obtain trustful result at any values of $\mu$ one
have to use transparent and free of any restrictions on $\mu$
procedure,
which would allow to perform calculations
with arbitrary nonabelian background gauge fields.

It was shown at zero chemical potential in \cite{ni,ni1,niemi}
that Chern-Simons term in odd dimensions
is connected with chiral anomaly in even dimensions by
trace identities.
As we'll see below generalization of the
trace identity on nonzero density is not trivial.
It connects chiral anomaly with Chern-Simons term,
which has $\mu$ and $T$ dependent coefficient.
We will see below that despite chemical potential and temperature
give rise to a coefficient in front of Chern-Simons term \cite{my}
they doesn't influence on chiral anomaly \cite{el10,chirm}.
Indeed, anomaly is  short distance phenomenon, which should not
be effected by medium (density and temperature) effects,
or more quantitatively,
so as  the anomaly has ultraviolet nature, temperature and chemical
potential should not give any ultraviolet effect since
distribution functions decrease exponentially with energy
in the ultraviolet limit.

The paper is organized as follows. In section 2
we briefly discuss introducing of the chemical
potential, chiral chemical potential and temperature to a theory.
Section 3 is devoted to qualitative consideration of
chiral anomaly in 2 and 4 dimensions. The rigorous proof
of density and temperature independence of axial anomaly
is presented in section 4. Also,
it is shown in 2-dimensional Schwinger model
that chiral anomaly is not influenced not only by chemical
potential $\mu$,
but also by Lagrange multiplier $\kappa$ at the constraint of
chiral charge conservation.
Section 5 is concerned to Chern--Simons term in even dimensions
and its reduction to odd dimension in high temperature limit.
In section 6 we obtain Chern--Simons term
in 3-dimensional theory at finite density and temperature
by use of a few different methods. In section 7 we evaluate
Chern--Simons term  in the presence of nonzero temperature and density
in 5-dimensional theory and generalize this result on
arbitrary nonabelian odd-dimensional theory.
Nonrelativistic consideration is presented in section 8.
In section 9 we generalize trace
identity on arbitrary density of background fermions on the basis of the
previous calculations.  Section 10 is devoted to concluding remarks.

\section{Chemical potential.}
As it is well-known, chemical potential can be introduced in a theory
as Lagrange multiplier at corresponding conservation laws.
In nonrelativistic physics this is  conservation of full number of
particles. In relativistic quantum field theory these are  conserving
charges.
The ground state energy
can be obtained by use of variational principle
\be
\langle \psi^{*} \hat{H} \psi \rangle =min
\ee
under charge conservation constraint for relativistic
equilibrium system
\be
\label{ch1}
\langle \psi^{*} \hat{Q} \psi \rangle ={{\rm Const}},
\ee
where $\hat{H}$ and $\hat{Q}$ are
hamiltonian and charge operators. Instead, we can use method of
undetermined Lagrange multipliers and seek absolute minimum of
expression
\be
\langle \psi^{*} (\hat{H}-\mu\hat{Q}) \psi \rangle ,
\ee
where $\mu$ is Lagrange multiplier.
Since $\hat{Q}$ commute with the hamiltonian
$\langle\hat{Q}\rangle$ is conserved.

On the other hand, we can impose
another constraint, which implies chiral charge conservation
\be
\langle \psi^{*} \hat{Q}_{5} \psi \rangle ={{\rm Const}},
\ee
or in Lagrange approach we have
\be
\langle \psi^{*} (\hat{H}-\kappa\hat{Q_{5}}) \psi \rangle = {\rm min},
\ee
where $\kappa$ arises as Lagrange multiplier at
$\langle\hat{Q}_{5}\rangle = Const$ constraint.
Thus, $\mu$ corresponds to nonvanishing fermion density
(number of particles minus number of antiparticles)
in background. Meanwhile, $\kappa$ is responsible for conserving
asymmetry in numbers of left and right handed background fermions.

It must be emphasized that the
formal addition of a chemical potential in the theory
looks like a simple gauge transformation
with the gauge function $\mu t$. However, it doesn't only shift the time
component
of a vector potential but also gives corresponding prescription for
handling Green's function poles.
The correct introduction of a chemical potential redefines
the ground state (Fermi energy),
which leads to a new spinor propagator with the correct
$\epsilon$-prescription for poles.
So, for the free spinor propagator we have
(see, for example, \cite{shur,chod})
\begin{eqnarray}
\label{sh}
G(p;\mu)=
\frac{\tilde{\not\! p}+m}
{(\tilde{p_{0}}+i\epsilon\sgn p_0 )^2-\vec{p}^2-m^2 },
\end{eqnarray}
where $\tilde{p}=(p_0 + \mu,\vec{p})$. Thus, when $\mu =0$ one at once
gets the usual $\epsilon$-prescription  because of the positivity of
$p_0\sgn p_0$.
In the presence of a  background Yang--Mills field we consequently
have for the Green function operator (in Minkovsky space)
\begin{eqnarray}
\label{gre}
\hat{G}=(\gamma \tilde{\pi}-m)\frac{1}{(\gamma\tilde{\pi})^2-m^2 + i \epsilon
(p_{0}+\mu)\sgn (p_{0})} ,
\end{eqnarray}
where $\tilde{\pi_{\nu}}=\pi_{\nu}+\mu\delta_{\nu 0}$ ,
$\pi_{\nu}=p_{\nu}-gA_{\nu}(x)$.

In Euclidian metric one has
\begin{eqnarray}
\label{euc}
G(p;\mu)=
\frac{\tilde{\not\! p}+m}
{\tilde{p_{0}}^2+\vec{p}^2+m^2 },
\end{eqnarray}
where $\tilde{p}=(p_0 + i\mu,\vec{p})$.

For temperature introduction we will use standard Matzubara approach
valid for systems in equilibrium. That is Euclidian generating
functional with temperature instead of time, and antiperiodic
conditions on fermion fields $\psi (0,\vec{x})=-\psi (\beta,\vec{x})$
and pereodic for boson ones $A (0,\vec{x})= A (\beta,\vec{x})$ .
Thus, for transfer to finite temperature case we will use
$$\int d^{D}x \rightarrow i\int_{0}^{\beta}dx_{0} \int d^{D-1} x,$$
$$\int \frac{d^{D}k}{(2\pi)^{D}} \rightarrow
\frac{i}{\beta}\sum_{n=-\infty}^{\infty} \int
\frac{d^{D-1}k}{(2\pi)^{D-1}},$$
together with  $p_{0}\rightarrow \omega_{n}=(2n+1)\pi/\beta$.
Here, the chemical potential also can be introduced by
adding it to a Matzubara frequency
$p_{0}\rightarrow \omega_{n}=(2n+1)\pi/\beta + i\mu$.

\section{Chiral anomaly. Qualitative consideration.}
First of all let us consider simple but rather intuitive than
rigorous derivation of axial anomaly \cite{nin}.
Let us start with 2--dimensional right-handed Weyl fermion theory
coupled to a uniform electric field $\dot A_{1} =E$ in
the temporal gauge.  The one component right-handed Weyl equation
for $\psi_{R}=1/2(1+\gamma_{5})\psi$ reads
\be
i \dot\psi_{R}(x)=(-i\partial_{x} -A^{1})\psi_{R}(x).
\ee
The dispersion law  is $\omega(P)=P$. Corresponding to the
classical equation of a charged particle in the presence of an
electric field where $\dot P =eE$, the acceleration of the
right-handed particles  in quantum theory is given by $\dot\omega=
\dot P= eE$. The creation rate of the right-handed particles per
unit time and unit length is determined by a charge of the
Fermi surface, which distinguishes the filled and unfilled
states. Let the quantization length be $L$; the density of states
per length $L$ is $L/2\pi$ and the rate of change of right-handed
particle number $N_{R}$ is
\be
\dot N_{R}= L^{-1}(L/2\pi)\dot\omega=(e/2\pi)E.
\ee
This particle creation is the axial anomaly. Consequently the
chiral charge $Q_{R}$ is not conserved and $\dot Q_{R}=\dot N_{R}=
(e/2\pi)E$. It follows from an analogous reasoning that the
annihilation rate of left-handed particles with the dispersion law
$\omega = -P$  is
\be
\dot N_{L}= - (e/2\pi)E.
\ee
Therefore the anomaly for the Dirac particles is
\be
\dot N_{R} - \dot N_{L} = (e/\pi)E,
\ee
which gives $\dot Q_{5}= (e/\pi) E$.

In 4 dimensions we first calculate the energy levels of the
right-handed Weyl fermion  in the presence of the applied
uniform magnetic field along the third direction given by
$$A^{2}=Hx^{1}\;\;\;\;\;{\rm and}\;\;\;\; A^{\mu}=0
\;\;\;\;\;{\rm otherwise}.$$
The solution to the equation for two-component right-handed
field $\psi_{R}$ of the form
\be
\label{e4}
\left[ i\partial /\partial t - ({\bf P}- e{\bf A}){\bf\sigma }\right]
\psi_{R}(x)=0
\ee
is expressed in terms of a solution of the auxiliary equation
\be
\label{e5}
\left[ i\partial /\partial t - ({\bf P}- e{\bf A}){\bf \sigma}\right]
\left[ i\partial /\partial t + ({\bf P}- e{\bf A}){\bf \sigma}\right]
\Phi=0
\ee
as
\be
\label{e6}
\psi_{R}=
\left[ i\partial /\partial t + ({\bf P}- e{\bf A}){\bf \sigma}\right]
\Phi.
\ee
From eq.(\ref{e5}) the energy and the $P_{2}, P_{3}$ eigenfunction
satisfies an equation of the harmonic oscillator type
$$
\left[ - (\partial/\partial^{1})^{2} + (eH)^{2}
(x^{1}+ P_{2}/eH)+(P_{3})^{2}+ eH\sigma\right]\Phi=\omega^{2}\Phi,
$$
where $\sigma=\pm 1$. The energy levels are given by the Landau
levels,
\be
\label{e44}
\omega(n,\sigma,P_{3})=\pm \left[eH(2n+1)+(P_{3})^{2}+eH\sigma
\right]^{1/2},\\  (n=0,1,2,...)\nonumber
\ee
except for the $n=0$ and $\sigma=-1$ mode where
\be
\omega(n=0,\sigma=-1,P_{3})=\pm P_{3}.
\ee
The eigenfunction takes the form
\be
\label{e9}
\Phi_{n\sigma}(x)=Nn\sigma \exp(-iP_{2}x^{2}-iP_{3}x^{3})\nonumber\\
\exp\left[-1/2eH(x_{1}+P_{2}/eH)^{2}\right]
H_{n}(x_{1}+P_{2}/eH)\xi(\sigma),
\ee
with $Nn\sigma$ as the normalization constant.
Here $\xi(\sigma)$ denotes the eigenfunctions of the Pauli
spin $\sigma_{3}$ which can be taken as
$\xi(1)={ 1\choose 0 }$ and $\xi(-1)={ 0\choose 1 }$.
The solution of (\ref{e5}) is obtained  by inserting (\ref{e9})
into (\ref{e6}). This leads to the relations
$$\psi_{R}^{(n+1,\sigma=-1)}=(N_{n+1,\sigma=-1}/N_{n,\sigma=1})
\psi_{R}^{(n,\sigma=1)}\;\;,\;\;\; n=0,1,...$$
and
$$ \psi_{R}^{(n=0,\sigma=-1)}=0, \;\; {\rm with}\;\; \omega = -P_{3}.$$
Thus the energy levels of $\psi_{R}$ are (\ref{e44}) and
\be
\omega(n=0,\sigma=-1,P_{3})=P_{3}.
\ee

Next a uniform electric field is turned on along the third
direction parallel to $H$. As for the zero mode $(n=0, \sigma=-1)$
the dispersion law is the same as that for 2 dimensions and
the creation rate of the particles is calculated in a similar manner.
It should be noticed that when $E$ is on adiabatically there is no
particle creation in the $n \! = 0$ modes. The density of
the state per length $L$ is $LeH/4\pi^{2}$ and the creation rate
is given by
\be
\dot{N_{R}}=L^{-1}(LeH/4\pi^{2})\dot{\omega}(n=0,\sigma=-1,P_{3})
=(e^2/4\pi^2)EH,
\ee
which equals to $\dot Q_{R}$.

For the left-handed fermions the annihilation rate of the
left-handed particles is
\be
\dot{N_{L}}=-(e^2/4\pi^2)EH,
\ee
which is $\dot Q_{L}$.

We then have for the Dirac field
\be
\dot Q_{5}=\dot{Q_{R}}-\dot{Q_{L}}=(e^2/2\pi^2)EH,
\ee
that is the chiral anomaly.

Now we can easily estimate influence of background density
on the anomaly in this approach.
From the above consideration we can see that the anomaly proportional to
the time derivative of the zero mode energy. Taking into account  that
nonzero fermion density influence just reduces to the
shift of the Landau levels on $\mu$ which
doesn't depend on time, we can conclude that finite density
doesn't influence on the chiral anomaly. The same arguments are
just for the Lagrange multiplier at the axial charge $\kappa$,
the only difference is that $\kappa$ makes shift for
left and right-handed fermions with opposite sign. So, until
$\mu$ ($\kappa$) is time independent it won't effect
on the chiral anomaly. We'd like to stress that here there were
made adiabatic approximation, when we turn on electric field.
So, this consideration is just a plot and it needs a strict proof.

\section{Chiral anomaly at fi\-nite tem\-pe\-ra\-ture and den\-si\-ty.}
\subsection{Two dimensions.}
Since anomaly term originates from the ultraviolete divergent part,
it is not expected to be changed by the temperature.
Indeed, it was shown in several papers
(see \cite{el10} and references therein).
Moreover, the same can be said about influence of
background fermion density that have been checked in the
works \cite{chirm}.

To clear understand the nature of anomaly $\mu$-independence
we'll first consider  the simplest case -- 2-dimensional QED
by use of Schwinger nonperturbative method \cite{shw}.
Thus, following for  Schwinger  one  writes
\be
J^{\mu}=-ig\tr\left[ \gamma^{\mu} G(x,x^{'})\exp\left(
-ig\int_{x^{'}}^{x}d\xi^{\mu}A_{\mu}(\xi)\right)\right]_{x^{'}
\rightarrow x}
\ee
where $G(x,x^{'})$ -- propagator satisfying following equation
\be
\gamma^{\mu}\left(\partial_{\mu}^{x}-igA_{\mu}(x)\right)G(x,x^{'})
=\delta (x-x^{'}).
\ee
Further we use  Shcwinger's  anzats
\be
G(x,x^{'})=G^{0}(x,x^{'})\exp\left[ ig(\phi (x)-\phi(x^{'}))\right]
\ee
where $G^{0}(x,x^{'})$ -- free propagator
$$\gamma^{\mu}\partial_{\mu}^{x}G^{0}(x,x^{'})=\delta (x-x^{'}).$$
Thus, for $\phi$ we can write $\gamma^{\mu}\partial_{\mu}\phi=
\gamma^{\mu}A_{\mu}$.  At finite density $G^{0}(x,x^{'})$
has the form
\be
G^{0}(x,x^{'})=\int\frac{d^{2}p}{(2\pi)^{2}} \e^{ip(x-x^{'})}
\frac{\not\! p}{p^{2}+i\varepsilon (p_{0}+\mu)\sgn p_{0}}=\nonumber\\
=-i\not\! \partial\Biggl[ \int\frac{d^{2}p}{(2\pi)^{2}} \e^{ip(x-x^{'})}
\frac{1}{p^{2}+i\varepsilon } -\nonumber\\
-2\int_{-\infty}^{+\infty}\frac{dp_{1}}{2\pi}
\int_{-\infty}^{+\infty}\frac{dp_{0}}{2\pi}
\theta (-\tilde{p}_{0}\sgn p_{0})\e^{ip(x-x^{'})}\Im m
\frac{1}{p^{2}+i\varepsilon }\Biggr].
\ee
So, beside the usual zero density part $\mu$--dependent one appears.
Further, we have to regularize current by use of symmetrical
limit $x\rightarrow x^{'}$. After some simple algebra it is clearly seen
that all $\mu$--dependent terms after taking off the limit  will
disappear. Thus, contribution to the current arises from the Shcwinger
part only, so
\be
J^{\mu} &=& i\frac{g^2}{2\pi}\left(\delta^{\mu\nu} -
\frac{\partial^{\mu}\partial^{\nu}}{\partial^{2}}\right)A_{\nu}\nonumber\\
J^{\mu}_{5} &=& i\frac{g^2}{2\pi}\left(\varepsilon^{\mu\nu} -
\varepsilon^{\mu\alpha}
\frac{\partial^{\alpha}\partial^{\nu}}{\partial^{2}}\right)A_{\nu},
\ee
and we get usual anomaly in chiral current
\be
\partial_{\mu}J^{\mu}=0 \;\; , \;\;\;\;\;
\partial_{\mu}J_{5}^{\mu}=i\frac{g^{2}}{2\pi}\varepsilon^{\mu\nu}
\partial_{\mu}A_{\nu}=i\frac{g^{2}}{4\pi}\*^{*}F.
\ee

It is natural  to introduce Lagrange multiplier $\kappa$
at corresponding constraint to support
the conservation of the $Q^{5}$--charge, i.e.
the difference of left and right fermion densities $Q_{L}-Q_{R}$.
Since $\kappa$ and $\mu$ are Lagrange multipliers at
corresponding conservation laws
they, in principle,  have to influence some way
on symmetry violation by a quantum corrections, i.e. on anomalies.
However, the rather amazing situation occurs.
The demand of chiral charge conservation
(instead of the usual charge conservation)
on the quantum level doesn't influence chiral anomaly.
Really, in 2-dimensions  introduction of Lagrange multiplier $\kappa$
at the chiral charge conservation gives the term
$\kappa\bar\psi \gamma^{5}\gamma^{0}\psi=
\kappa\bar\psi \gamma^{1}\psi$ in lagrangian.
So,  $\kappa$ affects in the same way as $\mu$, i.e.
$\kappa$ doesn't influence chiral anomaly
(it is also seen in direct calculations, which are similar
to presented above for the case with $\mu$).
That could be explained due to ultraviolet nature
of the chiral anomaly, while $\kappa$ ($\mu$) doesn't introduce
new divergences in the theory.

From the calculations  it is clearly seen the principle difference
of chiral anomaly and Chern-Simons. The ultraviolet regulator -- $P$-exponent
gives rise to the anomaly, but (as we'll see below) doesn't
influence on Chern-Simons. Thus, it is natural, that the anomaly doesn't
depend on $\mu$, $\kappa$ and $T$ because
it has ultraviolet regularization
nature, while
neither density nor temperature does influence on
ultraviolet behavior of the theory. The general and clear proof
of axial anomaly temperature independence in any even dimension
will be presented in
section 9 on the basis of the trace identities.


\subsection{Four dimensions.}
In \cite{chirm} direct calculations of axial anomaly
at finite temperature and den\-si\-ty in 4-di\-men\-sio\-nal gauge theory
were performed by use of
imaginary and real time formalism by Fujikawa method \cite{fu}.
Here we  present the derivation of the axial anomaly  by use of
the elegant Fujikawa procedure.
Considering a system of fermions and gauge bosons in thermodynamical
equilibrium at temperature $T=\beta^{-1}$ and nonzero  chemical potential
$\mu$ in the imaginary time formalism one reads the generating functional
of correlation functions
\be
Z[J_{\nu},\eta,\bar\eta]=\int {\cal D}A {\cal D}c {\cal D}\bar c
{\cal D} \psi {\cal D} \bar \psi \nonumber\\ \exp \left[
\int_{0}^{\beta}d \tau\int d^{3}x \left({\cal L}(\vec{x},\tau)
+ J_{\nu}A^{\nu}+\bar\psi\eta+\psi\bar\eta\right)\right],
\ee
where
$${\cal L}(\vec{x},\tau)={\cal L}_{\psi}+{\cal L}_{YM}+
{\cal L}_{c}+{\cal L}_{GF}$$
represents the effective Lagrangian
density of the $SU(N)$ Yang--Mills field $A=(A_{\rho}^{j})$
coupled to  fermion fields $\psi=(\psi_{\alpha}^{a})$,
$\bar\psi=(\bar\psi_{\alpha}^{a})$ and to Faddeev--Popov ghost fields
$c=(c_{\alpha})$, $\bar c=(\bar c_{\alpha})$.
$\eta=(\eta_{\alpha}^{a})$, $\bar\eta=(\bar\eta_{\alpha}^{a})$
and $J=(J_{\rho}^{j})$ are external sources.
$A_{\rho}^{j}$, $J_{\rho}^{j}$, $c_{\alpha}$, $\bar c_{\alpha}$
are periodic in $\tau$ with period $\beta$, while
$\psi_{\alpha}^{a}$, $\bar\psi_{\alpha}^{a}$,
$\eta_{\alpha}^{a}$, $\bar\eta_{\alpha}^{a}$  are antiperiodic.
Upper latin indices and lower greek ones indicate flavor and
$SU(N)$ internal (color) indices respectively, and $j=1,...,N^{2}-1$,
the number of standard $SU(N)$ generators $(T^{j})$.
${\cal L}_{YM}$ and ${\cal L}_{c}$ are standard Lagrangian densities
for Yang--Mills bosons and ghosts, while ${\cal L}_{GF}$
describes gauge fixing.
On the other hand, one has
$${\cal L}_{\psi}=\sum_{a=1}^{N_{f}}\bar\psi^{a}
(i\not\!\! D^{a}_{T,\mu}-m^{a})\psi^{a}$$
(lower color indices being also
implicitly contracted), with $N_{f}$ the number of flavors,
$$i\not\!\!\! D^{a}_{T,\mu}=i\not\!\!\! D_{T} +\mu^{a}\gamma_{0}$$
and
$$\not\!\! D_{T}= i\gamma^{0}(\partial /\partial \tau +A_{4})-
\gamma^{k}(\partial /\partial x^{k} +A_{k}),$$
where
$iA_{\rho}=gT^{j}A^{j}_{\rho}$ and the Wick rotation has been performed
in the imaginary time formalism ($x_{0}\rightarrow\tau=ix_{0}$,
$A_{0}\rightarrow A_{4}=-iA_{0}$) so that $\not\!\!\! D_{T}$
becomes Hermitian. It is considered a chemical potential $\mu^{a}$
for each flavor (there is no flavor mixing).

Following Fujikawa \cite{fu} we are interested in the chiral transformation
of the fermion fields:
$$\psi_{\alpha}^{a}\rightarrow
\exp [i\delta (\vec{x}, \tau)\gamma_{5}]\psi_{\alpha}^{a}\;,\;\;\;
\bar\psi_{\alpha}^{a}\rightarrow \bar\psi_{\alpha}^{a}
\exp [i\delta (\vec{x}, \tau)\gamma_{5}],$$
which produces a change in
the fermion measure ${\cal D} \psi {\cal D} \bar \psi
\rightarrow C {\cal D} \psi {\cal D} \bar \psi$, giving rise to
the anomaly factor in the chiral current conservation law, $C$, which
is the direct finite temperature and density extension of the
zero temperature and density factor appearing in \cite{fu}.
The chemical potential term is invariant under the above
chiral transformation. Then the only possible finite temperature and density
effects must be contained in $C$. In order to display them,
let us expand
$$\psi^{a}(\vec{x},\tau)=\sum_{n}a_{n}
\phi_{n}^{a}(\vec{x},\tau)\;\;,\;\;\;\;
\bar\psi^{a}(\vec{x},\tau)=\sum_{n}\bar b_{n}
{\phi_{n}^{a}}^{+}(\vec{x},\tau),$$
$a_{n}, \bar b_{n}$ being
elements of the Grassmann algebra. On the other hand,
$\phi_{n}^{a}(\vec{x},\tau)$, which is antiperiodic in $\tau$,
is an eigenfunction of the Hermitian operator
$\not \!\!\! D^{a}_{T,\mu}=\not \!\!\! D_{T} +\mu^{a}\gamma_{0}$, i.e.
$\not \!\!\! D^{a}_{T,\mu}\phi_{n}^{a}=\lambda_{n}\phi_{n}^{a}$,
($\lambda_{n}$ being real and color indices being omitted)
and it fulfills
$$\int_{0}^{\beta}d\tau\int d^{3}x {\phi_{n}^{a}}^{+}
\phi_{r}^{a}=\delta_{nr}.$$
Moreover, it can be Fourier--expanded as
\be
\label{al32}
\phi_{n}^{a}(\vec{x},\tau)=\frac{1}{\beta}\sum_{j=-\infty}^{\infty}
\int d^{3}k \e^{i\omega_{n}\tau} \e^{-i\vec{k}\vec{x}}
\phi_{j}^{a}(\vec{k}), \\ \omega_{n}=\frac{(2n+1)\pi}{\beta}.\nonumber
\ee
Then, the measure ${\cal D} \psi {\cal D} \bar \psi$ becomes
$\prod_{n}da_{n}\prod_{m}d\bar b_{m}$ and by extending
directly the zero temperature and density calculations \cite{fu},
one finds
\be
C=\exp\left[-2i\int_{0}^{\beta}d\tau\int d^{3}x \delta(\vec{x},\tau)
a(\vec{x},\tau)\right]
\ee
with
\be
a(\vec{x},\tau)=\sum_{a=1}^{N_{f}}\sum_{n}{\phi_{n}^{a}}^{+}(\vec{x},\tau)
\gamma_{5}\phi_{n}^{a}(\vec{x},\tau).
\ee
The finite temperature and density anomaly $a(\vec{x},\tau)$
can be regularized by extending again Fujikawa's trick as
\be
a(\vec{x},\tau)=\lim_{M\rightarrow\infty}
\sum_{a=1}^{N_{f}}\sum_{n}{\phi_{n}^{a}}^{+}(\vec{x},\tau)
\gamma_{5}\exp\left[ -M^{-2}(\not\!\!\! D_{T,\mu}^{a})^{2} \right]
\phi_{n}^{a}(\vec{x},\tau),
\ee
and by changing the basis vectors to "plane waves" with (\ref{al32}).
We remark that $\not\!\!\! D_{T,\mu}^{a}$ is equivalent to
$\not\!\!\! D_{T}$ provided that, in the latter, one replaces
$A_{4}$ by $A_{4}-i\mu^{a}$.
This replacement leaves $F_{\rho\nu}$ invariant
($F_{\rho\nu}=\partial_{\rho}A_{\nu}-\partial_{\nu}A_{\rho}+
[A_{\rho},A_{\nu}]$): notice that $\mu^{a}$ is constant
and $[\mu^{a},T^{j}]=0$ for $a=1...N_{f}$, $j=1...N^{2}-1$.
One obtains
\be
\label{al36}
a(\vec{x},\tau)=N_{f}\lim_{M\rightarrow\infty}
\tr\left(\gamma_{5}\left\{[\gamma^{\rho},\gamma^{\nu}]F_{\rho\nu}
\right\}^{2}\right)\frac{1}{8M^{2}}\frac{1}{\beta}\nonumber\\
\sum_{n=-\infty}^{\infty}\int\frac{d^3 k}{(2\pi)^{3}}
\exp\left[-\left( \frac{\omega_{n}^{2}+\vec{k}^{2}}{M^{2}}\right)  \right],
\ee
where the trace $\tr$ runs over both internal and $\gamma$ matrices
indices.

The infinite series on the right hand side of (\ref{al36}) displays
what is, quite likely, the most important difference between the
actual finite temperature and density case and the zero temperature
and density one treated in \cite{fu}. We recall the following formula
valid for any $M$:
\be
\label{al37}
\frac{1}{\beta}\sum_{n=-\infty}^{\infty}\exp\left[
-\frac{\pi^{2}}{M^{2}\beta^{2}}(2n+1)^{2}\right]=
\int_{-\infty}^{\infty}\frac{dk^{0}}{2\pi}\exp\left[
-\left(\frac{k^{0}}{M}\right)^{2}\right].
\ee
A simple derivation of (\ref{al37}) can be found in \cite{al17}.

By using (\ref{al37}) in (\ref{al36}) and taking the trace over
the $\gamma$ matrices, we arrive at the final formula:
\be
a(\vec{x},\tau)=-\frac{N_{f}}{16\pi^{2}}\tr\left(
\frac{1}{2}\epsilon^{\sigma\nu\rho\lambda}
F_{\rho\lambda}F_{\sigma\nu}\right),
\ee
where the trace now runs only over internal indices.
We can see that there are no finite temperature and density
corrections to the chiral anomaly, as we have expected
from the previous considerations.


\section{Chern-Simons in even dimensional theory}
It will be natural to introduce in lagrangian classical conservation
law --- the conservation of the $Q^{5}$--charge, i.e.
the difference of left and right fermion densities $Q_{L}-Q_{R}$.
Thus the lagrangian with constraint on $Q_5$ has the form
\begin{equation}
\label{37}
{\cal L}=\frac{1}{8}\tr FF  + \bar\psi\left(
i\hat\partial -g\hat A +i\kappa \gamma^{0} \gamma^{5}
\right) \psi
\end{equation}
Thus, if we will deal with such a lagrangian we must get theory in
which $Q^{5}$--charge is conserved. To get effective action
only background field dependent
we have to take over $d\bar\psi\,d\psi$ integration. There are two
ways to do it:  one can  calculate straight forward
by use of perturbation theory and get the effective action,
another one is proper time method.
Certainly, we have to take into account that
at high temperatures dimensional reduction takes place.
Thus, for example, vacuum polarization tensor in reduced ---
3--dimensional theory  can be written as
\be
\Pi_{ij}(p^{2})=\left( g_{ij}p^{2} -p_{i}p_{j} \right) \Pi^{(1)}(p^{2})
+ i e_{ijk}p^{k}\Pi^{(2)}(p^{2})+p_{i}p_{j}\Pi^{(3)}(p^{2}).
\ee
The part of the vacuum polarization tensor containing Levi-Chivita
tensor $i e_{ijk}p^{k}\Pi^{(2)}(p^{2})$ gives rise to Chern-Simons term.

It is convenient to rewrite lagrangian
in more appropriate form by use of projection operators
\begin{equation}
{\cal L}=\frac{1}{2}\tr FF  + \bar\psi_{L}\left(
i\hat\partial -g\hat A +i\kappa \gamma^{0}
\right) \psi_{L}+\bar\psi_{R}\left(
i\hat\partial -g\hat A -i\kappa \gamma^{0}
\right) \psi_{R}
\end{equation}
where we have used $I=P_{+}+P_{-}\,$, $\,\gamma^{5}=P_{+}-P_{-}\,$,
$\,P_{+}=\frac{1+\gamma^{5}}{2}\,$, $\,P_{-}=\frac{1-\gamma^{5}}{2}$.

So, now we can evaluate $J_{L}$ and $J_{R}$ separately. One can easily see
that the lagrangian we have got is absolutely analogous to finite
temperature and density lagrangian with left(right)-handed fermions
which was considered  in \cite{redl85} by use of perturbative expansion.

Thus we can immediately write the answer  for $J_{L}$ and $J_{R}$ currents.
\begin{equation}
J^{\mu}_{(L/R)}= (\pm)\frac{\kappa}{4\pi}\beta W[A]
\end{equation}
where $ W[A]$ --Chern--Simons term. And consequently for full current
and chiral current we'll get correspondingly
\begin{equation}
J^{\mu}= 0
\end{equation}
\begin{equation}
J^{\mu}_{5}= 2\frac{\kappa}{4\pi}\beta W[A]
\end{equation}


It is also possible to obtain Chern-Simons  at zero temperature
for $\kappa\not = 0$ with clear physical sense (see, for example, \cite{rub}
where chiral fermions are considered at finite density
and \cite{nin} where Weyl particles are considered).
In 2-dimensional Schwinger model there is chiral anomaly
\be
\label{anom}
\partial_{\mu} J_{\mu}^{5}=-\frac{1}{\pi}\varepsilon_{\mu\nu}F^{\mu\nu}.
\ee
It could be derived by using the picture of energy levels crossing,
see for example \cite{nin,shif}.
Here, we will exploit this method for consideration of Chern-Simons term.
Thus we will consider Schwinger model (\ref{37}) on a ring
with periodic for $A_{\mu}$ and
antiperiodic for $\psi$ boundary conditions
\be
\label{15}
A (x=-L/2 , t)=A (x=L/2 , t) ,\nonumber\\
\psi (x=-L/2 , t) =-\psi (x=L/2 , t).
\ee
Thus, fields $A$ and $\psi$ could be expanded in Fourie modes
$\exp (ikx 2\pi/L)$ for bosons and $\exp (i[k+1/2]x 2\pi/L)$
for fermions. The lagrangian (\ref{37}) is invariant under local gauge
transformations:
$$\psi \rightarrow \psi \e^{i\alpha(x,t)}\;,\;\;\;\;
A_{\mu} \rightarrow A_{\mu} + \partial_{\mu} \alpha (x,t).$$
It is easily seen that due to local gauge transformations,
we can put all modes of $A_{1}$ to be zero except for the zero-mode.
Thus,we can consider $A_{1}$ to be $x$-independent.
There exists another type of gauge transformations
(large gauge transformations)
$$ \alpha = \frac{2\pi}{L} nx,$$
where $n$ is  an integer number. Nevertheless, this gauge is not periodic,
it satisfies condition (\ref{15}). Really,
$\partial\alpha/\partial x={{\rm Const}}$  and
$\partial\alpha/\partial t=0$,  thus periodicity of $A_{\mu}$ is
conserved, the same is also true for $\psi$. So, we can consider
the model on the circle $[0, 2\pi/L]$.
Further, we use adiabatic
approximation, putting that $A_{1}$ independent on time
(a slight time dependence we will turn on later),
and that $A_{0}=0$.  This adiabatic approximation is quite
natural from the physical point of view, see for example elegant
consideration by Shifman \cite{shif}.
We now calculate the number density of real left(right) fermions
$n_{L/R}[A_{1}]$ and fermionic energy density $\varepsilon_{L/R}[A_{1}]$,
assuming that number density $n_{L/R}$ at $\dot A_{1}=0$
is fixed. Note that system with fixed $n_{L/R}$ can be prepared by
inserting fermions into the box, which is initially empty.

It is straightforward to calculate the fermionic spectrum at $A_{1}\not =0$,
\be
E_{L/R}=\frac{2\pi}{L}(k\mp N_{CS})~~~,~~~~~k=0,\pm 1, \pm 2,...~~,
\ee
where
\be
N_{CS}=\frac{1}{2\pi}\int A_{1} dx^{1}
\ee
is the Chern-Simons number  in (1+1) dimensions. As the gauge field changes
from zero to some fixed $A_{1}$, $[N_{CS}]$ levels of left handed fermions
cross zero from above and the same number of right handed fermionic levels
cross zero from below.  This means that $[N_{CS}]$ left handed fermions fill
the negative energy levels in the Dirac sea, see Fig.1., and the same number
of right handed fermions leave it. We would like to stress, that
in this physical clear picture it is essential use of the adiabatic
approximation. The number densities for
left(right) handed fermions are
\be
\label{anom1}
n_{L/R}[A_{1}]= n^{0}_{L/R} \mp n_{CS} + O(L^{-1}),
\ee
where $n_{CS}=N_{CS}/L$ is  the  average Chern-Simons density.

\unitlength=1.00mm
\special{em:linewidth 0.4pt}
\linethickness{0.4pt}
\begin{picture}(145.00,105.00)
\put(10.00,15.00){\vector(0,1){90.00}}
\put(10.00,50.00){\vector(1,0){60.00}}
\put(10.00,50.00){\line(2,-1){50.00}}
\put(60.00,30.00){\line(-2,1){50.00}}
\put(10.00,60.00){\line(2,-1){50.00}}
\put(60.00,40.00){\line(-2,1){50.00}}
\put(10.00,70.00){\line(2,-1){50.00}}
\put(60.00,50.00){\line(-2,1){50.00}}
\put(10.00,80.00){\line(2,-1){49.00}}
\put(59.00,60.00){\line(-2,1){49.00}}
\put(10.00,90.00){\line(2,-1){49.00}}
\put(59.00,70.00){\line(-2,1){49.00}}
\put(59.00,75.00){\line(-2,1){38.00}}
\put(33.00,93.00){\line(2,-1){26.00}}
\put(59.00,20.00){\line(-2,1){49.00}}
\put(10.00,40.00){\line(2,-1){37.00}}
\put(10.00,35.00){\line(2,-1){29.00}}
\put(66.00,43.00){\makebox(0,0)[cc]{A}}
\put(3.00,98.00){\makebox(0,0)[cc]{ {E(left)} }}
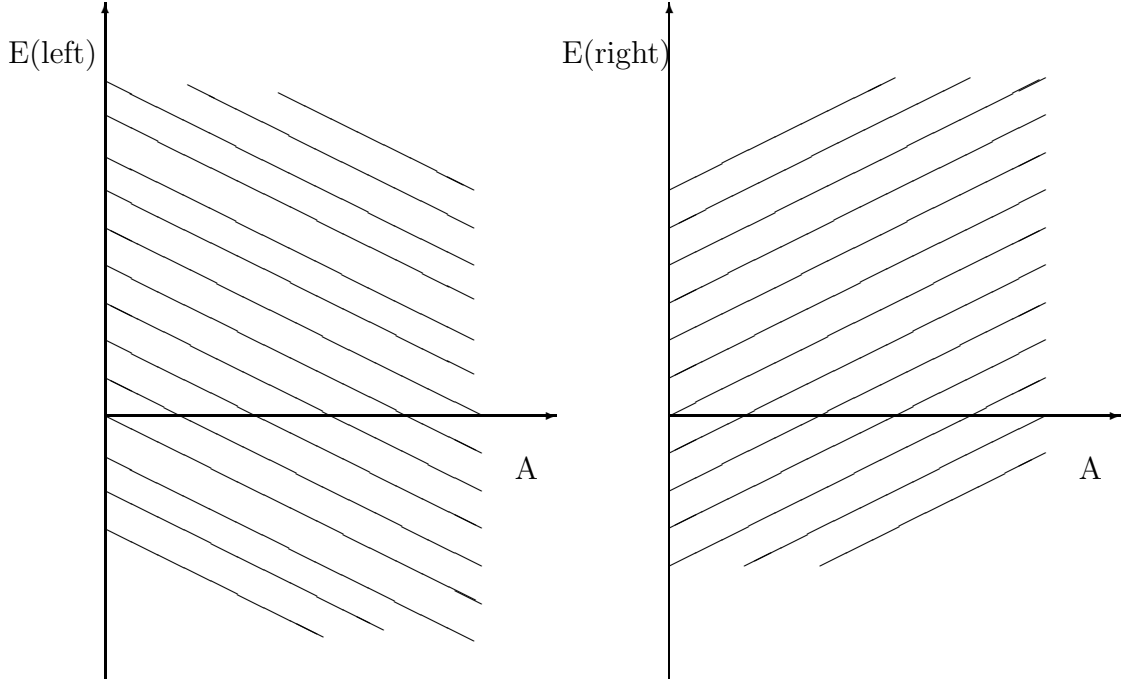
\put(3.00,5.00){\makebox(0,0)[cl]{Fig.1. Behavior of left(right)
handed fermionic levels.}}
\put(85.00,15.00){\vector(0,1){90.00}} \put(85.00,50.00){\vector(1,0){60.00}}
\put(141.00,43.00){\makebox(0,0)[cc]{A}}
\put(78.00,98.00){\makebox(0,0)[cc]{ {E(right)} }}
\put(85.00,50.00){\line(2,1){50.00}}
\put(135.00,70.00){\line(-2,-1){50.00}}
\put(85.00,40.00){\line(2,1){50.00}}
\put(135.00,60.00){\line(-2,-1){50.00}}
\put(85.00,30.00){\line(2,1){50.00}}
\put(135.00,50.00){\line(-2,-1){40.00}}
\put(105.00,30.00){\line(2,1){30.00}}
\put(135.00,80.00){\line(-2,-1){50.00}}
\put(85.00,60.00){\line(2,1){50.00}}
\put(135.00,90.00){\line(-2,-1){50.00}}
\put(85.00,70.00){\line(2,1){50.00}}
\put(125.00,95.00){\line(-2,-1){40.00}}
\put(85.00,80.00){\line(2,1){30.00}}
\end{picture}

Note that equation (\ref{anom1}) is essentially the integral form of the
anomaly equation (\ref{anom}). The average energy density of real fermions
is
\be
\varepsilon_{L/R}=\frac{2}{L}\sum_{[N_{CS}]+1}^{N^{0}_{L/R}/2}\varepsilon_{k}=
\frac{\pi}{2}(n^{0}_{CS}\mp n_{CS} )^{2} + O(L^{-2}).
\ee

We can introduce  chemical potential for left(right) handed fermions
in a standard way
\be
\mu_{L/R}=\frac{\partial \varepsilon_{L/R}}{\partial n_{L/R}}
\ee
and we obtain
\be
\mu_{L/R}=\pi (n_{L/R} \mp n_{CS}).
\ee
Introducing  the standard Legendre transform
\be
\tilde{E}_{L/R}[\mu_{L/R},A_{1}]=E_{L/R}\mp \mu_{L/R} N_{L/R}
\ee
we find
\be
\Delta E = (\mu_{R}  - \mu_{L}) N_{CS},
\ee
so, for the case when $\mu_{R}=\mu_{L}=\mu$ we'll get energy unchanged.
On the other hand, for chiral fermions \cite{rub} sign of term $\mu_{R}$
will change and
\be
\Delta E =  -2 \mu N_{CS}.
\ee
If we impose conservation of the left and  right handed fermions
(with Lagrangian multiple $\kappa$) instead of separate conservation
of left (right) fermions, we'll get
\be
\Delta E =  -2 \kappa N_{CS}.
\ee
Thus, the same result arises as for chiral fermions at finite density,
so as for usual fermions under conservation of chiral charge.
One should notice that here there were used two approximations.
The first one is time--independence of $A_{0}$, the second is
adiabatic approximation. Nevertheless, this consideration is valuable
due to construction of clear physical picture of the phenomenon.


The Chern-Simons term appearance  in even dimensional theory could
be shown in simple and clear way. The only thing we need for it is
temperature and density independence of chiral anomaly
(see previous sections).
From the definition  one has
\be
\frac{\partial I_{eff}}{\partial\kappa} =
\int d^{D}x\langle J_{5}^{0}\rangle.
\ee
Since axial anomaly doesn't depend on $\kappa$,
effective action  contains the term proportional to anomalous
$Q_{5}$ charge with $\kappa$ as a coefficient. The same is for a
chiral theory,  there  effective action contains the term
proportional to anomalous
$Q$ charge with $\mu$ as coefficient, see for example \cite{redl85,rub,rut}.
So, we have
\be
\label{z}
\Delta I_{eff} = -\kappa\int dx_{0} W[A]
\ee
in conventional gauge theory and
\be
\label{z1}
\Delta I_{eff}^{\rm chiral} =-\mu\int dx_{0} W[A]
\ee
in the chiral theory. Here $W[A]$ -- Chern-Simons term.
Thus we get Chern-Simons with Lagrange multiplier as a coefficient.

It is well-known that at nonzero temperature
in $\beta\rightarrow 0$ limit
the dimensional reduction effect occurs.
So, extra $t$-dependence of Chern-Simons term in (\ref{z}) disappears
and Chern-Simons can be treated as
a mass term in 3-dimensional theory with $i\kappa /T$ coefficient
(the same for chiral theory with $\mu$ see \cite{redl85}).
For anomalous parts of effective action we have
\be
\label{m3}
\Delta I_{eff} = -i\kappa\beta W[A] \;\;, \;\;\;\;\;\;
\Delta I_{eff}^{\rm chiral} =-i\mu\beta  W[A]
\ee
in conventional and chiral gauge theories correspondingly.
The only problem arises in a treating Chern-Simons as
a mass term is that  the coefficient is imaginary,
see discussions on the theme in \cite{redl85,rut}.
The other problem is that the coefficient is not the integer
function, see discussions in conclusion.
One can notice, that results (\ref{z}),(\ref{z1}) and (\ref{m3})
hold in arbitrary even dimension.
Let us stress, that we don't need any complicated calculations
to obtain (\ref{z}--\ref{m3}). The only thing we need is
the knowledge of  chiral anomaly independence on $\mu$, $\kappa$
and $\beta$.

This result also can be derived by use of the proper--time method.
Chiral current reads as follows
\be
\label{59}
J^{\mu}_{5}=-ig\tr\left[\gamma^{5}\gamma^{\mu}
G(x,x^{'})\right]=
-ig\tr\left[\gamma^{5}\gamma^{\mu}
\frac{1}{i\not\!\partial -g\not\! A +\gamma^{5}\gamma^{0}\kappa}\right]=
\nonumber\\=
-ig\tr\left[\gamma^{\mu}
\frac{1}{i\not\!\partial -g\gamma^{5}\not\! A +\gamma^{0}\kappa}\right].
\ee
The propagator has the following form
\be
G(x,x^{'})=\left[i\not\!\partial -g\gamma^{5}\not\! A \right] (-i)
\int_{-\infty}^{0}d\tau U(x,x^{'};\tau)
\ee
where $U(x,x^{'};\tau)$ -- evolution operator in a proper time.
The propagator after substitution of the evolution operator
can be rewritten as
\be
G(x,x^{'})=\exp\left( -ig\gamma^{5}\int_{x^{'}}^{x}d\zeta_{\mu}A^{\mu}\right)
\int\frac{d^{4}p}{(2\pi)^{4}}\e^{ip(x-x^{'})}G(p)
\ee
where $G(p)$ has the form
\be
G(p)=-ia_{d}\int_{-\infty}^{0}\frac{d\tau}{\tau^{d/2}}
\exp\Bigl(-\frac{1}{2}\tr \ln\left[\frac{i}{4\tau}\ch ( gF\tau )
\right]-\nonumber\\ -ip (gF)^{-1}\th (gF\tau) p \Bigr)
\left[ \gamma^{5}\gamma^{\alpha}\left( \th (gF\tau) \right)^{\alpha\nu}
p^{\nu} -\not\! p \right]\nonumber\\
\exp\left(i\frac{g}{2}\gamma^{5}
\sigma_{\mu\nu}F^{\mu\nu}\tau\right),
\ee
where $a_{d}=\e^{i\pi d/4}/(2\pi)^{d/2}$.
Substituting expression for the propagator in (\ref{59}),
we will get for the chiral current
\be
J^{\mu}_{5}=g a_{d}\int_{-\infty}^{0}\frac{d\tau}{\tau^{d/2}}
\exp\left(-\frac{1}{2}\tr \ln\left[\frac{i}{4\tau}\ch ( gF\tau )
\right] \right)\nonumber\\ \int\frac{d^{4}p}{(2\pi)^{4}}
\exp\left(-ip (gF)^{-1}\th (gF\tau) p \right) p^{\nu}
\nonumber\\ \tr\Bigl[
\left( \gamma^{\mu}\gamma^{5}\gamma^{\alpha}
\left[ \th (gF\tau) \right]^{\alpha\nu}
 -\gamma^{\mu}\gamma^{\alpha}g^{\alpha\nu} \right)
\exp\left(i\frac{g}{2}\gamma^{5}
\sigma_{\mu\nu}F^{\mu\nu}\tau\right)
\Bigr].
\ee
Taking into account that $p$--integration is at finite density and
temperature, i.e. integral in $p_{0}$ is changed on sum,
and extracting the Levi-Chivita tensor containing part
(its really simple, if one take traces in covariant form)
we obtain
\be
J^{\mu}_{5}=\frac{g^{2}}{8\pi^{2}}\beta\int\frac{d\tau}{\tau^{2}}
\sum_{m=1}^{\infty}(-1)^{m}m\exp\left(i\frac{\beta^{2}m^{2}}{4\tau}\right)
\sh (\beta m\kappa)\;^{*}F^{\mu 0}
\ee
To regulate this expression we use dimensional regularization, which
can be expressed in terms of generalized Riemann zeta functions.
Also, we take high temperature limit, i.e. $\beta\rightarrow0$,
and finally get
\be
J^{\mu}_{5}=i\kappa \frac{g^{2}}{2\pi^{2}}\;^{*}F^{\mu 0}.
\ee

\section{Chern-Simons in 3-dimensional theory.}
\subsection{Constant magnetic field.}

Let us first consider a (2+1) dimensional  abelian theory
in the external constant magnetic field.
We will evaluate
fermion density by performing
the direct summation over Landau levels.
As a starting point, we will use the formulae for fermion number at
finite density and temperature \cite{ni}
\be
\label{num}
N=- \frac{1}{2}\sum_{n}\sgn (\lambda_{n}) +
\sum_{n}\Bigl[ \frac{\theta (\lambda_{n})}{\exp
(-\beta(\mu -\lambda_{n}))+1} -\nonumber\\ -
\frac{\theta (-\lambda_{n})}{\exp (-\beta(\lambda_{n} -\mu ))+1}\Bigr]
=\nonumber\\=
\frac{1}{2}\sum_{n}\th \frac{1}{2}\beta(\mu -\lambda_{n})
\stackrel{\beta\rightarrow\infty}{\longrightarrow}
\frac{1}{2}\sum_{n}\sgn (\mu -\lambda_{n}) .
\ee
Landau levels in the constant magnetic field have the form \cite{lan}
\be
\lambda_{0} = -m\sgn (eB)~~~,~~~~~~\lambda_{n}=\pm
\sqrt{2n|eB|+m^2}
\ee
where n=1,2, ...
It is also necessary to take into account in (\ref{num}) the degeneracy
of  Landau levels.
Namely, the number of degenerate states for each Landau level
is $|eB|/2\pi$ per unit  area.
Even now we can see that only zero modes
(because of $\sgn (eB)$)
could contribute to the parity odd quantity.
So, for zero temperature, by using the identity
\be
\sgn (a - b) +\sgn (a + b)=2\sgn (a) \theta (|a| -|b|),\nonumber
\ee
one gets for zero modes
\be
\frac{|eB|}{4\pi}\sgn \left(\mu + m\sgn (eB)\right)=
\frac{|eB|}{4\pi}\sgn (\mu) \theta (|\mu | -|m|) + \nonumber\\ +
\frac{|eB|}{4\pi}\sgn (eB)\sgn (m) \theta (|m | -|\mu|),
\ee
and for nonzero modes
\be
\frac{1}{2}\frac{|eB|}{2\pi}\sum_{n=1}^{\infty}\sgn (\mu - \sqrt{2n|eB|+m^2})
+\sgn (\mu + \sqrt{2n|eB|+m^2})=\nonumber\\
=\frac{|eB|}{2\pi}\sgn (\mu) \sum_{n=1}^{\infty}\theta (|\mu | -
\sqrt{2n|eB|+m^2}).
\ee
Combining contributions of all modes we get for fermion  density
\be
\label{rho}
\rho=\frac{|eB|}{2\pi}\sgn (\mu) \sum_{n=1}^{\infty}
\theta\left(|\mu | - \sqrt{2n|eB|+m^2}\right) +\nonumber\\ +
\frac{1}{2}\frac{|eB|}{2\pi}\sgn (\mu) \theta (|\mu | -|m|)
+\frac{1}{2}\frac{eB}{2\pi}\sgn (m) \theta (|m | -|\mu|)=\nonumber\\
=\frac{|eB|}{2\pi}\sgn (\mu)\left( {\rm Int}\left[\frac{\mu^2- m^2}{2 |eB|}
\right]+\frac{1}{2}\right)
\theta (|\mu|-|m|)+\nonumber\\ + \frac{eB}{4\pi}\sgn (m)
\theta (|m | -|\mu|).
\ee
Here we see that zero modes contribute as to parity odd so as to parity
even part, while  nonzero modes contribute to the parity even part only
(note that under parity transformation $B\rightarrow -B$).
Thus, fermion  density  contains as Chern-Simons part,
so as parity even part.
At finite temperature it is also possible to get Chern-Simons.
Substituting zero modes into (\ref{num}) one gets
\be
N_{0}=\frac{|eB|}{2\pi}\frac{1}{2}\th\left[ \frac{1}{2}\beta
\left( \mu + m\sgn (eB) \right)  \right] =\nonumber\\=
\frac{|eB|}{4\pi}\left[ \frac{\sh (\beta\mu)}{\ch (\beta\mu)+\ch (\beta m)}
+\sgn (eB) \frac{\sh (\beta m)}{\ch (\beta\mu)+\ch (\beta m)}
\right],
\ee
so, excluding parity odd part, one gets for Chern-Simons at finite temperature
and density
\be
N_{CS}=
\frac{eB}{4\pi}\frac{\sh (\beta m)}{\ch (\beta\mu)+\ch (\beta m)}=
\frac{eB}{4\pi}\th (\beta m)\frac{1}{1+\ch (\beta\mu)/\ch (\beta m)}.
\ee
So, the result coincides with result
for Chern-Simons term coefficient by Niemi and Semenoff \cite{ni1}
obtained for small $\mu$
$$\left[ \th \frac{1}{2}\beta(m-\mu)+\th \frac{1}{2}\beta(m+\mu)\right].$$
It is obvious the limit to zero temperature.
The lack of this method is
that it works only for abelian and constant field case.

This result at zero temperature can be obtained by use of
Schwinger proper--time
method. Consider  (2+1) dimensional theory in the abelian case and choose
background field in the form
$$ A^{\mu}=\frac{1}{2}x_{\nu}F^{\nu\mu} ,
\;\;\;\;\;\; F^{\nu\mu}={\rm Const}. $$
To obtain the  Chern-Simons term in this case, it is necessary to consider
the background current
$$ \langle J^{\mu}\rangle =  \frac{\delta S_{eff}}{\delta A_{\mu}}  $$
rather than the effective action itself. This is because the Chern-Simons
term formally vanishes for such the choice of
$A^{\mu}$  but its variation with respect to $A^{\mu}$
produces a nonvanishing current.
So, consider
\begin{equation}
\langle J^{\mu}\rangle= -i g \tr\left[\gamma^{\mu}G(x,x^{'})
\right]_{x\rightarrow x^{'}}
\end{equation}
where
\begin{equation}
G(x,x^{'})=\exp\left(-ig\int_{x^{'}}^{x}d\zeta_{\mu}A^{\mu}(\zeta)\right)
\langle x | \hat G | x^{'}\rangle .
\end{equation}
Let us rewrite  Green function (\ref{gre}) in a more appropriate form
\begin{eqnarray}
\hat{G}=(\gamma \tilde{\pi}-m)
\Bigl[ \frac{\theta ((p_{0}+\mu)\sgn (p_{0}))}
{(\gamma\tilde{\pi})^2-m^2 + i \epsilon }+
\frac{\theta (-(p_{0}+\mu)\sgn (p_{0}))}
{(\gamma\tilde{\pi})^2-m^2 - i \epsilon }
\Bigr] .
\end{eqnarray}
Now, we  use the well known integral representation  of denominators
$$\frac{1}{\alpha \pm i0}=\mp i\int_{0}^{\infty}ds\e^{\pm i\alpha s},$$
which corresponds to introducing the ''proper--time'' $s$ into the
calculation
of the Eulier--Hei\-sen\-berg lagrangian by the Schwinger method \cite{schwin}.
We obtain
\begin{eqnarray}
\hat{G}=i(\gamma \tilde{\pi}-m)\int_{0}^{\infty} ds
\Bigl[ -  \exp\left( i s \left[
(\gamma\tilde{\pi})^2-m^2 + i \epsilon \right]\right)\nonumber\\
\theta ((p_{0}+\mu)\sgn (p_{0}))+ \nonumber\\ +
\exp\Bigl( -i s \left[
(\gamma\tilde{\pi})^2-m^2 - i \epsilon \right]\Bigr)
\theta (-(p_{0}+\mu)\sgn (p_{0}))
\Bigr] .
\end{eqnarray}
For simplicity, we restrict ourselves only to the magnetic field case,
where $A_{0}=0, [\tilde\pi_{0},\tilde\pi_{\mu}]=0 $. Then we easily
can factorize the time dependent part of Green function
\be
G(x,x^{'})=\int \frac{d^3 p}{(2\pi)^3}\hat G \e^{ip(x-x^{'})}=\nonumber\\
=\int \frac{d^2 p}{(2\pi)^2}\hat G_{\vec{x}}
\e^{i\vec{p}(\vec{x}-\vec{x}^{'})}
\int \frac{d p_0}{2\pi}\hat G_{x_0} \e^{ip_0 (x_0 -x_0^{'})}.
\ee
By using the obvious relation
\begin{equation}
\label{yr}
(\gamma\tilde{\pi})^2=(p_0 +\mu)^2 -\vec{\pi}^2 +\frac{1}{2}g
\sigma_{\mu\nu}F^{\mu\nu}
\end{equation}
one gets
\begin{eqnarray}
G(x,x^{'})|_{x\rightarrow x^{'}}= -i\int\frac{dp_{0}}{2\pi}
\frac{d^2 p}{(2\pi)^2}(\gamma \tilde{\pi}-m)\int_{0}^{\infty} ds\nonumber\\
\Biggl[
\e^{is(\tilde{p}_{0}^{2}-m^2)}\e^{-is\vec{\pi}^2}
\e^{isg\sigma F/2}
-
\theta (-(p_{0}+\mu)\sgn (p_{0}))\nonumber\\ \Bigl(
\e^{is(\tilde{p}_{0}^{2}-m^2)}\e^{-is\vec{\pi}^2}
\e^{isg\sigma F/2}   +
\e^{-is(\tilde{p}_{0}^{2}-m^2)}\e^{is\vec{\pi}^2}
\e^{-isg\sigma F/2}\Bigr)
\Biggr] .
\end{eqnarray}
Here the first term corresponds to the usual $\mu$--independent case and
there are two additional $\mu$--dependent terms.
In the calculation of the current the following trace arises:
\be
\tr \left[ \gamma^{\mu} (\gamma\tilde{\pi}-m)
\e^{isg\sigma F/2} \right]=
2\pi^{\nu}g^{\nu\mu}\cos ( g|^{*}\! F|s )+\nonumber\\
+2\frac{\pi^{\nu}F^{\nu\mu}}{|^{*}\! F|}\sin ( g|^{*}\! F|s ) -
2im\frac{^{*}\! F^{\mu}}{|^{*}\! F|}\sin ( g|^{*}\! F|s),\nonumber
\ee
where $\*^{*}\! F^{\mu}=\varepsilon^{\mu\alpha\beta}F_{\alpha\beta}/2$
and $ |^{*}\! F|=\sqrt{B^2-E^2} $.
Since we are interested in  calculation of the
parity odd part (Chern-Simons term)
it is enough to consider only
terms proportional to the dual strength tensor $\*^{*}\! F^{\mu}$.
On the other hand the term $2\pi^{\nu}g^{\nu\mu}\cos \left( g|^{*}\! F|s\right)$
at $\nu=0$ (see expression for the trace,
we take in mind that here there is only magnetic field)
also gives nonzero contribution to
the current $J^{0}$ \cite{tolpa}
\be
J^{0}_{\rm even}=g\frac{|gB|}{2\pi}\left( {\rm Int}\left[
\frac{\mu^{2}-m^{2}}{2|gB|}
\right]+\frac{1}{2}\right) \theta (|\mu | -|m|).
\ee
This part of current is parity invariant because under parity
$B\rightarrow -B$.
It is clear that this parity even object
does  contribute
neither to the parity anomaly nor to the mass of the
gauge field.
Moreover,  this term has been obtained \cite{tolpa}  in the pure
magnetic background and scalar magnetic field occurs
in the argument's denominator of the cumbersome function --
integer part. So, the parity even term seems to be
''noncovariantizable'', i.e. it can't be converted in covariant
form in effective action.
For a pity,  in  papers \cite{tolpa} charge density
consisting of both parity odd and parity even parts
is dubbed Chern-Simons, what leads to misunderstanding.
The main goal of this article is to explore the parity anomalous
topological Chern-Simons term in the effective action at finite density.
So, just the
term proportional to the dual strength tensor $\*^{*}\! F^{\mu}$ will
be considered.
The relevant part of the current reads
\begin{eqnarray}
J^{\mu}_{CS}=\frac{g}{2\pi} \int dp_{0}\int \frac{d^2 p}{(2\pi)^2}
\int_{0}^{\infty} ds
\frac{2im^{*}\! F^{\mu}}{|^{*}\! F|}\sin \left( g|^{*}\! F|s\right)\nonumber\\
\Biggl[
\e^{is(\tilde{p}_{0}^{2}-m^2)}\e^{-is\vec{\pi}^2}
-
\theta (-(p_{0}+\mu)\sgn (p_{0})) \nonumber\\ \left(
\e^{is(\tilde{p}_{0}^{2}-m^2)}\e^{-is\vec{\pi}^2} -
\e^{-is(\tilde{p}_{0}^{2}-m^2)}\e^{is\vec{\pi}^2}
\right)\Biggr] .
\end{eqnarray}
Evaluating integral over spatial momentum we derive
\begin{eqnarray}
J^{\mu}_{CS}=\frac{g^2}{4\pi^2} m \*^{*}\! F^{\mu}
\int_{-\infty}^{+\infty} dp_{0} \int_{0}^{\infty} ds
\Biggl[
\e^{is(\tilde{p}_{0}^{2}-m^2)}-\nonumber\\
-\theta (-\tilde{p}_{0}\sgn (p_{0}))\left(
\e^{is(\tilde{p}_{0}^{2}-m^2)}+
\e^{-is(\tilde{p}_{0}^{2}-m^2)m}\right)\Biggr] .
\end{eqnarray}
Thus, we have got besides the usual Chern--Simons part \cite{redl}, also
the $\mu$--dependent one.
It is easy to calculate it by use of the formula
$$ \int_{0}^{\infty}ds \e^{is(x^2-m^2)}=\pi\left(
\delta (x^2-m^2) +\frac{i}{\pi} {\cal{P}} \frac{1}{x^2-m^2}\right) $$
and we  get eventually
\begin{eqnarray}
\label{fab}
J^{\mu}_{CS}&=&\frac{m}{|m|}\frac{g^2}{4\pi}\*^{*}\! F^{\mu}
\bigl[ 1 - \theta (-(m+\mu)\sgn(m))
-\theta (-(m-\mu)\sgn(m))\bigr]\nonumber\\
&=&\frac{m}{|m|}\theta (m^2 -\mu^2 )\frac{g^2}{4\pi}\*^{*}\! F^{\mu}.
\end{eqnarray}

Let us now discuss the non-abelian case. Then  $A^{\mu}=T_{a} A_{a}^{\mu}$
and current reads
$$\langle J_{a}^{\mu}\rangle= -i g \tr\left[\gamma^{\mu} T_{a}
G(x,x^{'})\right]_{x\rightarrow x^{'}} .$$
It is well--known \cite{redl,red11}
that there exist only two types of the constant background fields.
The first is the  ''abelian'' type
(it is easy to see that the self--interaction
$f^{abc}A_{b}^{\mu}A_{c}^{\nu}$
disappears under that choice of the background field)
\begin{equation}
\label{ab}
A_{a}^{\mu}=\eta_{a}\frac{1}{2}x_{\nu}F^{\nu\mu},
\end{equation}
where $\eta_{a}$ is an arbitrary constant
vector in the color space, $F^{\nu\mu}={\rm Const}$.
The second  is the pure ''non--abelian'' type
\begin{equation}
\label{nab}
A^{\mu}={\rm Const}.
\end{equation}
Here the derivative terms (abelian part) vanish from the strength tensor
and it contains only the self--interaction part
$F^{\mu\nu}_{a}=gf^{abc}A_{b}^{\mu}A_{c}^{\nu}$.
It is clear that to catch  abelian part of the Chern-Simons term
we should consider the background field (\ref{ab}),
whereas for the non--abelian (derivative noncontaining,
cubic in $A$) part
we have to use the case (\ref{nab}).

Calculations in the ''abelian'' case reduces to the previous analysis,
except the trivial adding  of the color indices in the formula (\ref{fab}):
\begin{eqnarray}
\label{finab}
J^{\mu}_{a}
=\frac{m}{|m|}\theta (m^2 -\mu^2 )
\frac{g^2}{4\pi}\*^{*}F^{\mu}_{a} .
\end{eqnarray}
In the case (\ref{nab})  all calculations are similar. The only
difference is that the origin of term $\sigma_{\mu\nu}F^{\mu\nu}$
in (\ref{yr}) is not the linearity $A$ in $x$ (as in abelian case) but
the pure non--abelian   $A^{\mu}={\rm Const}$. Here  term
$\sigma_{\mu\nu}F^{\mu\nu}$ in (\ref{yr}) becomes quadratic in $A$
and we have
\begin{eqnarray}
\label{finnab}
J^{\mu}_{a}
=\frac{m}{|m|}\theta (m^2 -\mu^2 )
\frac{g^3}{4\pi}\varepsilon^{\mu\alpha\beta}
\tr\left[ T_{a}A^{\alpha}A^{\beta} \right] .
\end{eqnarray}
Combining formulas (\ref{finab}) and (\ref{finnab}) and integrating
over field $A^{\mu}_{a}$  we obtain eventually
\begin{equation}
S^{{\rm CS}}_{eff}=\frac{m}{|m|}\theta (m^2 -\mu^2 ) \pi W[A] ,
\end{equation}
where $W[A]$ is the Chern-Simons term
$$W[A]=\frac{g^2}{8\pi^2}\int d^{3}x \varepsilon^{\mu\nu\alpha}
\tr \left( F_{\mu\nu}A_{\alpha}-
\frac{2}{3}gA_{\mu}A_{\nu}A_{\alpha}\right) .$$
It may seem that covariant notation is rather artificial.
However, it helps us to extract Levi-Chivita tensor containing
part of action, i.e. parity anomalous Chern-Simons.

\subsection{Arbitrary external field.}
One can see that the methods we have used above for calculation of
Chern-Simons term
are noncovariant. Indeed, both of them use the constant
magnetic background.
Therefore, here we will use completely covariant
approach, which allows an arbitrary initial field
configuration and nonabelian fields.
We will employ the perturbative expansion
at once in the nonabelian case.

Let us first consider nonabelian 3--dimensional gauge theory.
The only graphs whose P-odd parts  contribute to the
parity anomalous Chern-Simons term are shown in Fig.2.

\unitlength=1.00mm
\special{em:linewidth 0.4pt}
\linethickness{0.4pt}
\begin{picture}(80.00,50.00)
\put(25.00,35.00){\circle{10.20}}
\put(65.00,35.00){\circle{10.00}}
\put(69.00,32.00){\line(1,0){1.00}}
\put(70.00,32.00){\line(0,-1){1.00}}
\put(70.00,31.00){\line(1,0){1.00}}
\put(71.00,31.00){\line(0,-1){1.00}}
\put(71.00,30.00){\line(1,0){1.00}}
\put(72.00,30.00){\line(0,-1){1.00}}
\put(72.00,29.00){\line(1,0){1.00}}
\put(73.00,29.00){\line(0,-1){1.00}}
\put(73.00,28.00){\line(1,0){1.00}}
\put(74.00,28.00){\line(0,-1){1.00}}
\put(61.00,32.00){\line(-1,0){1.00}}
\put(60.00,32.00){\line(0,-1){1.00}}
\put(60.00,31.00){\line(-1,0){1.00}}
\put(59.00,31.00){\line(0,-1){1.00}}
\put(59.00,30.00){\line(-1,0){1.00}}
\put(58.00,30.00){\line(0,-1){1.00}}
\put(58.00,29.00){\line(-1,0){1.00}}
\put(57.00,29.00){\line(0,-1){1.00}}
\put(57.00,28.00){\line(-1,0){1.00}}
\put(56.00,28.00){\line(0,-1){1.00}}
\put(22.00,20.00){\makebox(0,0)[lb]{(a)}}
\put(63.00,20.00){\makebox(0,0)[lb]{(b)}}
\put(2.00,10.00){\makebox(0,0)[lb]{
{Fig.2  Graphs whose P-odd parts contribute to}}}
\put(2.00,4.00){\makebox(0,0)[lb]{
{Chern-Simons term in nonabelian 3D gauge theory}}}
\put(30.50,35.50){\oval(1.00,1.00)[t]}
\put(31.50,35.50){\oval(1.00,1.00)[t]}
\put(32.50,35.50){\oval(1.00,1.00)[t]}
\put(33.50,35.50){\oval(1.00,1.00)[t]}
\put(34.50,35.50){\oval(1.00,1.00)[t]}
\put(35.50,35.50){\oval(1.00,1.00)[t]}
\put(14.50,35.50){\oval(1.00,1.00)[t]}
\put(15.50,35.50){\oval(1.00,1.00)[t]}
\put(16.50,35.50){\oval(1.00,1.00)[t]}
\put(17.50,35.50){\oval(1.00,1.00)[t]}
\put(18.50,35.50){\oval(1.00,1.00)[t]}
\put(19.50,35.50){\oval(1.00,1.00)[t]}
\put(65.50,40.50){\oval(1.00,1.00)[lt]}
\put(65.50,41.50){\oval(1.00,1.00)[rt]}
\put(65.50,42.50){\oval(1.00,1.00)[lt]}
\put(65.50,43.50){\oval(1.00,1.00)[rt]}
\put(65.50,44.50){\oval(1.00,1.00)[lt]}
\put(65.50,45.50){\oval(1.00,1.00)[rt]}
\end{picture}

Thus, the part of effective action containing the Chern-Simons term looks as
\be
\label{eff}
I^{CS}_{eff} =
\frac{1}{2}\int_{x}  A_{\mu}(x)\int_{p}\e^{-ixp}A_{\nu}(p)
\Pi^{\mu\nu}(p)\nonumber\\ +
\frac{1}{3}\int_{x}  A_{\mu}(x)\int_{p,r}\e^{-ix(p+r)}
A_{\nu}(p)A_{\alpha}(r)\Pi^{\mu\nu\alpha}(p,r),
\ee
where polarization operator and vertice have a standard form
\be
\Pi^{\mu\nu}(p) &=&g^2 \int_{k} \tr \left[ \gamma^{\mu}S(p+k;\mu)
\gamma^{\nu}S(k;\mu)\right] \nonumber\\
\Pi^{\mu\nu\alpha}(p,r) &=& g^3
\int_{k}\tr \left[ \gamma^{\mu}S(p+r+k;\mu)
\gamma^{\nu}S(r+k;\mu)\gamma^{\alpha}S(k;\mu).
\right],
\ee
here, under  integration we understand
$$\int_{x}=i\int_{0}^{\beta}dx_{0}
\int d\vec{x}\;\;\;{\rm and}\;\;\;\;
\int_{k}=\frac{i}{\beta}\sum_{n=-\infty}^{\infty}
\int \frac{d\vec{k}}{(2\pi)^{2}}.$$
First consider the second order term (Fig.2, graph (a)).
It is well-known that the only object giving us the
possibility to construct $P$ and $T$ odd form in action
is Levi-Chivita tensor\footnote{In three dimensions it arises as
a trace  of three $\gamma$--matrices (Pauli matrices)}. Thus,
we will drop all terms noncontaining Levi-Chivita tensor.
Signal for the mass generation (Chern-Simons term) is
$\Pi^{\mu\nu}(p^{2}=0)\not =0$. So  we  get
\be
\Pi^{\mu\nu}=g^2 \int_{k} ( -i2m e^{\mu\nu\alpha} p_{\alpha} )
\frac{1}{(\tilde{k}^2 +m^2 )^2}.
\ee
After some simple algebra one  obtains
\be
\Pi^{\mu\nu}=-i2mg^2e^{\mu\nu\alpha} p_{\alpha}
\frac{i}{\beta}\sum_{n=-\infty}^{\infty}\int
\frac{d^{2}k}{(2\pi)^2}
\frac{1}{(\tilde{k}^2 +m^2 )^2} =\nonumber\\
=-i2mg^2e^{\mu\nu\alpha} p_{\alpha}
\frac{i}{\beta}\sum_{n=-\infty}^{\infty}\frac{i}{4\pi}
\frac{1}{\omega_{n}^{2}+m^2},
\ee
where $\omega_{n}=(2n+1)\pi / \beta +i\mu$. Performing
summation we get
\be
\label{6}
\Pi^{\mu\nu}=i\frac{g^2}{4\pi}e^{\mu\nu\alpha} p_{\alpha}
\th (\beta m)\frac{1}{1+\ch (\beta\mu)/\ch (\beta m)}
\ee
It is easily seen that at $\beta\rightarrow\infty$ limit we'll
get zero temperature result \cite{my}
\be
\Pi^{\mu\nu}=i\frac{m}{|m|}\frac{g^2}{4\pi}e^{\mu\nu\alpha} p_{\alpha}
\theta (m^2 -\mu^2 ) .
\ee
In the same manner handling the third order contribution (Fig.2b)
one gets
\be
\Pi^{\mu\nu\alpha} &=& -2g^3 i e^{\mu\nu\alpha}
\frac{i}{\beta}\sum_{n=-\infty}^{\infty}\int\frac{d^2 k}{(2\pi)^2}
\frac{m(\tilde{k}^2+m^2)}{ (\tilde{k}^2 +m^2 )^3}
=\nonumber\\&=&-i2mg^3  e^{\mu\nu\alpha}
\frac{i}{\beta}\sum_{n=-\infty}^{\infty}\int\frac{d^2 k}{(2\pi)^2}
\frac{1}{ ( \tilde{k}^2 +m^2 )^2 }
\ee
and  further all calculations are identical to the second order
\be
\label{8}
\Pi^{\mu\nu\alpha}
=i\frac{g^3}{4\pi}e^{\mu\nu\alpha}
\th (\beta m)\frac{1}{1+\ch (\beta\mu)/\ch (\beta m)}.
\ee
Substituting (\ref{6}), (\ref{8}) in the effective action
(\ref{eff}) we get eventually
\be
I^{CS}_{eff} =\th (\beta m)\frac{1}{1+\ch (\beta\mu)/\ch (\beta m)}
\frac{g^2}{8\pi}
\int d^{3}x e^{\mu\nu\alpha}\nonumber\\  \tr\left(
A_{\mu}\partial_{\nu}A_{\alpha}-
\frac{2}{3}g A_{\mu}A_{\nu}A_{\alpha}\right).
\ee
Thus, we get Chern-Simons term with temperature and
density dependent coefficient.

\section{Chern-Simons term in arbitrary odd dimension.}

Let's now consider 5--dimensional gauge theory.
Here the Levi-Chivita ten\-sor is 5--dimen\-sional
$e^{\mu\nu\alpha\beta\gamma}$
and the rele\-vant graphs are shown in Fig.3.

\unitlength=1.00mm
\special{em:linewidth 0.4pt}
\linethickness{0.4pt}
\begin{picture}(120.00,50.00)
\put(25.00,35.00){\circle{10.20}}
\put(65.00,35.00){\circle{10.00}}
\put(105.00,35.00){\circle{10.00}}
\put(25.50,40.50){\oval(1.00,1.00)[lt]}
\put(25.50,41.50){\oval(1.00,1.00)[rt]}
\put(25.50,42.50){\oval(1.00,1.00)[lt]}
\put(25.50,43.50){\oval(1.00,1.00)[rt]}
\put(25.50,44.50){\oval(1.00,1.00)[lt]}
\put(25.50,45.50){\oval(1.00,1.00)[rt]}
\put(105.50,40.50){\oval(1.00,1.00)[lt]}
\put(105.50,41.50){\oval(1.00,1.00)[rt]}
\put(105.50,42.50){\oval(1.00,1.00)[lt]}
\put(105.50,43.50){\oval(1.00,1.00)[rt]}
\put(105.50,44.50){\oval(1.00,1.00)[lt]}
\put(105.50,45.50){\oval(1.00,1.00)[rt]}
\put(23.00,20.00){\makebox(0,0)[lb]{(a)}}
\put(63.00,20.00){\makebox(0,0)[lb]{(b)}}
\put(103.00,20.00){\makebox(0,0)[lb]{(c)}}
\put(3.00,10.00){\makebox(0,0)[lb]{
{Fig.3  Graphs whose P-odd parts contribute to }}}
\put(3.00,4.00){\makebox(0,0)[lb]{
{Chern-Simons term in nonabelian 5D theory}}}
\put(21.00,32.00){\line(-1,0){1.00}}
\put(20.00,32.00){\line(0,-1){1.00}}
\put(20.00,31.00){\line(-1,0){1.00}}
\put(19.00,31.00){\line(0,-1){1.00}}
\put(19.00,30.00){\line(-1,0){1.00}}
\put(18.00,30.00){\line(0,-1){1.00}}
\put(18.00,29.00){\line(-1,0){1.00}}
\put(17.00,29.00){\line(0,-1){1.00}}
\put(17.00,28.00){\line(-1,0){1.00}}
\put(16.00,28.00){\line(0,-1){1.00}}
\put(29.00,32.00){\line(1,0){1.00}}
\put(30.00,32.00){\line(0,-1){1.00}}
\put(30.00,31.00){\line(1,0){1.00}}
\put(31.00,31.00){\line(0,-1){1.00}}
\put(31.00,30.00){\line(1,0){1.00}}
\put(32.00,30.00){\line(0,-1){1.00}}
\put(32.00,29.00){\line(1,0){1.00}}
\put(33.00,29.00){\line(0,-1){1.00}}
\put(33.00,28.00){\line(1,0){1.00}}
\put(34.00,28.00){\line(0,-1){1.00}}
\put(61.00,32.00){\line(-1,0){1.00}}
\put(60.00,32.00){\line(0,-1){1.00}}
\put(60.00,31.00){\line(-1,0){1.00}}
\put(59.00,31.00){\line(0,-1){1.00}}
\put(59.00,30.00){\line(-1,0){1.00}}
\put(58.00,30.00){\line(0,-1){1.00}}
\put(58.00,29.00){\line(-1,0){1.00}}
\put(57.00,29.00){\line(0,-1){1.00}}
\put(57.00,28.00){\line(-1,0){1.00}}
\put(56.00,28.00){\line(0,-1){1.00}}
\put(101.00,32.00){\line(-1,0){1.00}}
\put(100.00,32.00){\line(0,-1){1.00}}
\put(100.00,31.00){\line(-1,0){1.00}}
\put(99.00,31.00){\line(0,-1){1.00}}
\put(99.00,30.00){\line(-1,0){1.00}}
\put(98.00,30.00){\line(0,-1){1.00}}
\put(98.00,29.00){\line(-1,0){1.00}}
\put(97.00,29.00){\line(0,-1){1.00}}
\put(97.00,28.00){\line(-1,0){1.00}}
\put(96.00,28.00){\line(0,-1){1.00}}
\put(74.00,43.00){\line(-1,0){1.00}}
\put(73.00,43.00){\line(0,-1){1.00}}
\put(73.00,42.00){\line(-1,0){1.00}}
\put(72.00,42.00){\line(0,-1){1.00}}
\put(72.00,41.00){\line(-1,0){1.00}}
\put(71.00,41.00){\line(0,-1){1.00}}
\put(71.00,40.00){\line(-1,0){1.00}}
\put(70.00,40.00){\line(0,-1){1.00}}
\put(70.00,39.00){\line(-1,0){1.00}}
\put(69.00,39.00){\line(0,-1){1.00}}
\put(115.00,42.00){\line(-1,0){1.00}}
\put(114.00,42.00){\line(0,-1){1.00}}
\put(114.00,41.00){\line(-1,0){1.00}}
\put(113.00,41.00){\line(0,-1){1.00}}
\put(113.00,40.00){\line(-1,0){1.00}}
\put(112.00,40.00){\line(0,-1){1.00}}
\put(112.00,39.00){\line(-1,0){1.00}}
\put(111.00,39.00){\line(0,-1){1.00}}
\put(111.00,38.00){\line(-1,0){1.00}}
\put(110.00,38.00){\line(0,-1){1.00}}
\put(56.00,43.00){\line(1,0){1.00}}
\put(57.00,43.00){\line(0,-1){1.00}}
\put(57.00,42.00){\line(1,0){1.00}}
\put(58.00,42.00){\line(0,-1){1.00}}
\put(58.00,41.00){\line(1,0){1.00}}
\put(59.00,41.00){\line(0,-1){1.00}}
\put(59.00,40.00){\line(1,0){1.00}}
\put(60.00,40.00){\line(0,-1){1.00}}
\put(60.00,39.00){\line(1,0){1.00}}
\put(61.00,39.00){\line(0,-1){1.00}}
\put(95.00,42.00){\line(1,0){1.00}}
\put(96.00,42.00){\line(0,-1){1.00}}
\put(96.00,41.00){\line(1,0){1.00}}
\put(97.00,41.00){\line(0,-1){1.00}}
\put(97.00,40.00){\line(1,0){1.00}}
\put(98.00,40.00){\line(0,-1){1.00}}
\put(98.00,39.00){\line(1,0){1.00}}
\put(99.00,39.00){\line(0,-1){1.00}}
\put(99.00,38.00){\line(1,0){1.00}}
\put(100.00,38.00){\line(0,-1){1.00}}
\put(69.00,32.00){\line(1,0){1.00}}
\put(70.00,32.00){\line(0,-1){1.00}}
\put(70.00,31.00){\line(1,0){1.00}}
\put(71.00,31.00){\line(0,-1){1.00}}
\put(71.00,30.00){\line(1,0){1.00}}
\put(72.00,30.00){\line(0,-1){1.00}}
\put(72.00,29.00){\line(1,0){1.00}}
\put(73.00,29.00){\line(0,-1){1.00}}
\put(73.00,28.00){\line(1,0){1.00}}
\put(74.00,28.00){\line(0,-1){1.00}}
\put(109.00,32.00){\line(1,0){1.00}}
\put(110.00,32.00){\line(0,-1){1.00}}
\put(110.00,31.00){\line(1,0){1.00}}
\put(111.00,31.00){\line(0,-1){1.00}}
\put(111.00,30.00){\line(1,0){1.00}}
\put(112.00,30.00){\line(0,-1){1.00}}
\put(112.00,29.00){\line(1,0){1.00}}
\put(113.00,29.00){\line(0,-1){1.00}}
\put(113.00,28.00){\line(1,0){1.00}}
\put(114.00,28.00){\line(0,-1){1.00}}
\end{picture}

The part of effective action containing Chern-Simons term reads
\be
\label{ef}
I^{CS}_{eff} =
\frac{1}{3}\int_{x}  A_{\mu}(x)\int_{p,r}\e^{-ix(p+r)}
A_{\nu}(p)A_{\alpha}(r)\Pi^{\mu\nu\alpha}(p,r)\nonumber\\
+
\frac{1}{4}\int_{x}  A_{\mu}(x)\int_{p,r}\e^{-ix(p+r+s)}
A_{\nu}(p)A_{\alpha}(r)A_{\beta}(s)\Pi^{\mu\nu\alpha\beta}(p,r,s)\nonumber\\
+
\frac{1}{5}\int_{x}  A_{\mu}(x)\int_{p,r}\e^{-ix(p+r+s+q)}
A_{\nu}(p)A_{\alpha}(r)A_{\beta}(s)A_{\gamma}(s)\nonumber\\
\times\Pi^{\mu\nu\alpha\beta\gamma}(p,r,s,q)
\ee
All calculations are similar to 3--dimensional case.
First consider third order contribution (Fig.3a)
\be
\Pi^{\mu\nu\alpha}(p,r) = g^3
\int_{k}\tr \left[ \gamma^{\mu}S(p+r+k;\mu)
\gamma^{\nu}S(r+k;\mu)\gamma^{\alpha}S(k;\mu)
\right].
\ee
Taking into account that trace of five $\gamma$-matrices in
5--dimensions is
$$
\tr\left[\gamma^{\mu}\gamma^{\nu}\gamma^{\alpha}\gamma^{\beta}\gamma^{\rho}
\right] = 4ie^{\mu\nu\alpha\beta\rho}, $$
we  extract the parity odd part of the vertice
\be
\Pi^{\mu\nu\alpha}=g^3 \frac{i}{\beta}\sum_{n=-\infty}^{\infty}
\int\frac{d^4 k}{(2\pi)^4}
( i4m e^{\mu\nu\alpha\beta\sigma} p_{\beta}r_{\sigma} )
\frac{1}{(\tilde{k}^2 +m^2 )^3},
\ee
or in more transparent way
\be
\Pi^{\mu\nu\alpha}&=& i4mg^3 e^{\mu\nu\alpha\beta\sigma}
p_{\alpha}r_{\sigma} \frac{i}{\beta} \sum_{n=-\infty}^{+\infty}
\int\frac{d^{4}k}{(2\pi)^4}
\frac{1}{(\omega_{n}^2 +\vec{k}^2 +m^2 )^3}=\nonumber\\&=&
i4mg^3 e^{\mu\nu\alpha\beta\sigma}
p_{\alpha}r_{\sigma} \frac{i}{\beta} \sum_{n=-\infty}^{+\infty}
\frac{-i}{64\pi^2} \frac{1}{\omega_{n}^2+m^2}.
\ee
Evaluating summation one comes to
\be
\label{p1}
\Pi^{\mu\nu\alpha}
=i\th (\beta m)\frac{1}{1+\ch (\beta\mu)/\ch (\beta m)}
\frac{g^3}{16\pi^2}e^{\mu\nu\alpha\beta\sigma}
p_{\alpha}r_{\sigma}.
\ee
In the same way operating graphs (b) and (c) (Fig.3) one will obtain
\be
\Pi^{\mu\nu\alpha\beta}
=i\th (\beta m)\frac{1}{1+\ch (\beta\mu)/\ch (\beta m)}
\frac{g^4}{8\pi^2}e^{\mu\nu\alpha\beta\sigma}s_{\sigma}
\ee
and
\be
\label{p2}
\Pi^{\mu\nu\alpha\beta\gamma}
=i\th (\beta m)\frac{1}{1+\ch (\beta\mu)/\ch (\beta m)}
\frac{g^5}{16\pi^2}e^{\mu\nu\alpha\beta\sigma}.
\ee
Substituting (\ref{p1}) --- (\ref{p2}) in the effective
action (\ref{ef}) we get the
final result for  Chern-Simons in 5--dimensional theory
\be
I^{CS}_{eff} =\th (\beta m)\frac{1}{1+\ch (\beta\mu)/\ch (\beta m)}
\frac{g^3}{48\pi^2}
\int_{x} e^{\mu\nu\alpha\beta\gamma} \nonumber\\
\tr\left(
A_{\mu}\partial_{\nu}A_{\alpha}\partial_{\beta}A_{\gamma}+
\frac{3}{2}g A_{\mu}A_{\nu}A_{\alpha}\partial_{\beta}A_{\gamma}+
\frac{3}{5}g^{2} A_{\mu}A_{\nu}A_{\alpha}A_{\beta}A_{\gamma}
\right).
\ee

It is remarkable that all parity odd contributions are finite
as in 3--dimensional so as in 5--dimensional cases.
Thus, all values in the effective action are renormalized in
a standard way, i.e. the renormalizations are
determined by conventional (parity even) parts of vertices.

From the above direct calculations  it is clearly seen
that the chemical potential and temperature dependent coefficient is the
same for all parity odd parts of diagrams and doesn't depend on
space dimension. So, the  influence
of finite density and temperature on Chern-Simons term generation is the same in
any odd dimension:
\begin{equation}
\label{kon}
I^{{\rm CS}}_{eff}=\th (\beta m)\frac{1}{1+\ch (\beta\mu)/\ch (\beta m)}
 \pi W[A] \stackrel{\beta\rightarrow\infty}{\longrightarrow}
\frac{m}{|m|}\theta (m^2 -\mu^2 )\pi W[A],
\end{equation}
where $W[A]$ is the Chern-Simons secondary characteristic class
in any odd dimension.
Since  only the lowest orders
of perturbative series contribute to Chern-Simons term at finite density and temperature
(the same situation
is well-known at zero density), the result obtained by using
formally perturbative technique appears to be nonperturbative.
Thus, the $\mu$ and $T$ --dependent Chern-Simons term coefficient
reveals the amazing property of universality.
Namely, it does depend on
neither dimension of the theory nor abelian or nonabelian gauge
theory is studied.

The arbitrariness of $\mu$ gives us the possibility
to see Chern-Simons coefficient behavior at any masses.
It is very interesting that  $\mu^2 = m^2$ is the
crucial point for Chern-Simons at zero temperature.
Indeed, it is clearly seen from (\ref{kon}) that when $\mu^2 < m^2$
$\mu$--influence disappears and we get the usual Chern-Simons term
$$I^{{\rm CS}}_{eff}= \pi W[A].$$
On the other hand when $\mu^2 > m^2$
the situation is absolutely different.
One can see that here the Chern-Simons term
disappears because of non--zero density of background fermions.
We would like to emphasize the
important massless case $m=0$ considered in many a papers,
see for example \cite{ni1,redl,jac}.
Here even negligible density or temperature,
which always take place in any
physical processes, leads to vanishing of the parity anomaly.
Let us stress again that we nowhere have used
any restrictions on $\mu$.
Thus we not only confirm result
of \cite{ni1} for Chern-Simons in $QED_{3}$ at small density,
but also expand it
on arbitrary $\mu$, nonabelian case and arbitrary odd dimension.


\section{Nonrelativistic consideration.}
Here, we will show that in nonrelativistic case there are
no Chern-Simons term, there is only pseudo Chern-Simons,
which is even under parity transformation.
It is also presented the possibility to get mixed Chern-Simons
term in nontrivial external field.

First, we would like to notice that
there are two approaches in fermion number definition.
The first one is (see for example \cite{soni})
\be
\langle Q\rangle_{\beta,\mu}=\sum_{n}\frac{1}{e^{\beta(\lambda_{n}-\mu)}+1},
\ee
and a normal ordering is performed at the given value  of the
chemical potential $\mu$.
(This normal ordering is suppressed here since it is
inessential to the present discussion.) The other definition
(see \cite{ni}) is related to the above by
\be
\langle Q\rangle_{\beta,\mu}=\langle N\rangle_{\beta,\mu}  +
\frac{1}{2}\zeta_{H}(0),
\ee
where $\zeta_{H}$ is the Riemann $\zeta$-function related to the
even part of the spectral density of the hamiltonian $H$
\be
\zeta_{H}(s)=\int_{0}^{\infty}d\lambda\left[\rho_{H}(\lambda)+
\rho_{H}(-\lambda)\right]\lambda^{-s}.
\ee
So, the difference in the definitions is given by a $\beta$ and $\mu$
independent constant, $\zeta_{H}(0)$.
Indeed, one can easily check that at the operator level, these two
definitions related as
\be
{\bf Q}={\bf N} + \frac{1}{2}\int dx\left\{\psi^{+}(x),\psi(x)\right\}\;\;,
\;\;\;\;{\bf N}=\frac{1}{2}\int dx\left[\psi^{+}(x),\psi(x)\right].
\ee

As we have seen above, the fermion number density has the following form
\be
N=\frac{1}{2}\sum_{n}\th \frac{1}{2}\beta(\mu -\lambda_{n})
\stackrel{\beta\rightarrow\infty}{\longrightarrow}
\frac{1}{2}\sum_{n}\sgn (\mu -\lambda_{n}) .
\ee
Landau levels in the relativistic case are
\be
\lambda_{0} = -m\sgn (eB)~~~,~~~~~~\lambda_{n}=\pm
\sqrt{2n|eB|+m^2}
\ee
where $n=1,2,...$ .
On the other hand, in the nonrelativistic case energy levels have the form
\be
\lambda_{n}=(n+\frac{1}{2})\Omega
\ee
where $\Omega=|eB|/m$ cyclotron frequency, $n=0,1,2,$...

As we have seen above in the relativistic case
fermion  density has the form
\be
N=
\frac{|eB|}{2\pi}\sgn (\mu)\left( {\rm Int}\left[\frac{\mu^2- m^2}{2 |eB|}
\right]+\frac{1}{2}\right)\theta (|\mu|-|m|)+\nonumber\\ +
\frac{eB}{4\pi}\sgn (m)\theta (|m | -|\mu|).
\ee
Thus we can see that in the relativistic case there is especial zero mode,
the only mode which contribute to parity odd part of fermion number.
In a contrast, in the nonrelativistic case there is no especial zero
mode, all modes contribute to the parity even part only.
Thus, we have at zero temperature
\be
Q=\frac{|eB|}{2\pi}\sum_{n}\theta\left(\mu-(n+\frac{1}{2})\Omega\right)
=\frac{|eB|}{2\pi} {\rm Int}\left[\frac{\mu m}{|eB|}+\frac{1}{2}\right].
\ee
One can see that fermion number in the nonrelativistic case
is parity even ($B\rightarrow -B$ under parity).
Therefore, it does not give rise  to the parity odd Chern-Simons term in
action. Instead of being variational derivative of the true Chern-Simons,
fermion number is derivative of the pseudo Chern-Simons \cite{schak}
\be
\langle Q\rangle =\frac{\delta}{\delta A_{0}} I_{pseudo CS}.
\ee

In the same manner it is possible to get fermion number with temperature
introduced. For example, such calculations were done in \cite{sal},
there was used another method. There the pseudo Chern-Simons term
coefficient has the form
\be
\Pi^{E}_{1}=\frac{1}{\pi}\sum_{n}\left(\exp\beta(\lambda_{n}-\mu)
+1\right)^{-1} -\nonumber\\ -\frac{1}{8\pi}\frac{\beta}{ml^{2}}
\sum_{n}(2n+1){\rm sech}^{2} \left(\frac{1}{2}\beta(\lambda_{n}-\mu))
\right).
\ee
It is clearly seen that this expression can be rewritten
in the way
\be
\Pi^{E}_{1}=\frac{1}{\pi}\sum_{n}
\Bigl[\frac{1}{\exp\beta(\lambda_{n}-\mu)+1}-\nonumber\\ -\frac{\beta\lambda_{n}}
{\exp\beta(\lambda_{n}-\mu)+\exp( -\beta(\lambda_{n}-\mu)) +2}
\Bigr].
\ee
After taking $\beta\rightarrow\infty$ limit one get
\be
\Pi^{E}_{1}=\frac{1}{\pi}\sum_{n}\theta (\mu - \lambda_{n}),
\ee
that coincides with the above  calculations.

Another paper is \cite{schak}. There were also considered
chemical potential influence on fermion number in nonrelativistic case.
In this section we treat a nonrelativistic electron gas confined to a
plane. We expect that some new qualitative features arise from the fact
that in this case the spin degree of freedom is not enslaved by the
dynamics. We continue to use a relativistic notation with
$\partial_\mu=(\partial_0, {\bf \nabla}$,
$\partial ^\mu =(\partial_0, -{\bf \nabla})$, where
${\bf \nabla}$ is the gradient operator,
and $A^\mu = (A^0, {\bf A})$.

Let us consider the Lagrangian
\begin{equation}
{\cal L} =\Psi ^{\dagger}(i\partial _{0}+\mu -H_{{\rm P}})\Psi +
      b \Psi ^{\dagger}\frac{\sigma ^{3}}{2}\Psi          \label{30}
\end{equation}
which governs the dynamics of the Pauli spinor field $\Psi$,
with Grassmann components $\psi _{\uparrow}$ and
$\psi _{\downarrow}$ describing the electrons with spin-$\uparrow$
and $\downarrow$. The role of the chemical potential $\mu$ and the
spin source $b$ is the same as in the previous calculation. The Pauli
Hamiltonian
\begin{equation}
H_{{\rm P}}=\frac{1}{2m}(i {\bf \nabla}
+e {\bf A})^{2}- g_0 \mu_{\rm B}
\frac{\sigma^3}{2} B + eA_{0},                              \label{31}
\end{equation}
with $\mu_{\rm B}=e/2m$ the Bohr magneton and $g_0$ the electron
$g$-factor,
contains a Zeeman term which couples the electron spins to the
background magnetic field. Usually this term is omitted. The reason is
that in realistic systems the $g$-factor is much larger than two, the
value for a free electron. In strong magnetic fields relevant to the
QHE the energy levels of spin-$\downarrow$ electrons are
too high and cannot be occupied; the system is spin
polarized, and the electron spin is irrelevant to the problem.
Setting again $A^{0}=A^{1}=0,\ A^{2}=B x^{1}$, one finds as eigenvalues
for $H_{{\rm P}}$
\begin{equation}
E_{n,\pm} = \frac{|eB|}{m} (n+{1 \over 2}) - \frac{eB}{m} S_\pm,
                                                         \label{nrll}
\end{equation}
with $S_\pm = \pm {1 \over 2}$ for spin-$\uparrow$ and spin-$\downarrow$
electrons, respectively. We note that in the nonrelativistic limit,
corresponding to taking $m \rightarrow + \infty$, the relativistic
Landau levels reduce to
\begin{equation}
E_{+n} \rightarrow {{\rm Const}} + \frac{|eB|}{m} (n+{1 \over 2}) -
\frac{eB}{2m},
\end{equation}
where we omitted the negative energy levels which have no meaning in
this limit. The main difference with (\ref{nrll}) stems from the fact
that there the spin degree of freedom is considered as an independent
quantity, not enslaved by the dynamics as is the case in the
relativistic problem.

The induced fermion number density and spin density may be obtained in a
similar calculation as in the preceding section. From the effective
action,
\begin{equation}
S_{{\rm eff}}=-i\tr\ln(i\partial _{0}-H_{P}+\mu +\frac{b}{2}\sigma
^{3}),
\end{equation}
one obtains
\be
{\cal L}_{{\rm eff}}=\frac{|eB |}{2\pi } \sum_{n=0}^{\infty }
\int \frac{dk_{0}}{2 \pi i} \Bigl[ \ln (k_{0}-E_{n,+}+\mu +\frac{b}{2})+\nonumber\\
+
\ln (k_{0}-E_{n,-}+\mu -\frac{b}{2})\Bigr].            \label{nreff}
\ee

The resulting value of the induced fermion number density is
\begin{equation}
\rho = \frac{|eB|}{2\pi} (N_+ + N_-),                       \label{39}
\end{equation}
with $N_\pm$ the number of filled Landau levels for spin-$\uparrow$ and
spin-$\downarrow$ electrons,
\begin{equation}
N_\pm = \left[ \frac{m \mu_\pm }{|eB|} + \frac{1}{2}\right] ,
                                                      \label{npm}
\end{equation}
and
\begin{equation}
\mu_\pm = \mu + \frac{eB}{m} S_\pm                     \label{chemical}
\end{equation}
their effective chemical potentials.
The square brackets denote again the integer-part function.
Implicit in this framework is the assumption that, just like in the
relativistic case, the chemical potential lies between two Landau levels.
The induced fermion number density (\ref{39}) is related to a
Chern-Simons term in the effective action, with a coefficient
\begin{equation}
\theta = {\rm  sgn}(eB) \frac{1}{2\pi} (N_+ + N_-).      \label{nrth}
\end{equation}
Because of the presence of the ${\rm  sgn}(eB)$ factor, which changes
sign under a parity transformation, this Chern-Simons term is invariant
under such transformations. The induced spin density turns out to be
independent of $N_\pm$, viz.\
\begin{equation}
s=\frac{eB }{4\pi}.                                       \label{41}
\end{equation}
This follows from the symmetry in the spectrum $E_{n+1,+} =
E_{n,-}$ $(eB>0)$, or $E_{n,+} = E_{n+1,-}$
$(eB<0)$. The magnetic moment, $M$ can be obtained
from (\ref{41}) by multiplying $s$ with twice the Bohr magneton,
$\mu_{\rm B}$. This leads to the text-book result for the magnetic spin
susceptibity $\chi_{\rm P}$
\begin{equation}
\chi_{\rm P} = \frac{\partial M}{\partial B} = \frac{e^2}{4 \pi m} = \
2 \mu_{\rm B}^2 \, \nu_{2D}(0),                        \label{textb}
\end{equation}
with $\nu_{2D}(0) = m/(2\pi)$ the density of states per spin degree of
freedom in two space dimensions.

At zero field, $\rho$ reduces to the standard fermion number density in
two space dimensions
$\rho \rightarrow \mu m/\pi = k_{\rm F}^2/(2\pi)$,
where $k_{{\rm F}}$ denotes the Fermi momentum. A single fluxon carries
according to (\ref{41}) a spin $S_\otimes = {1 \over 2}$ and, since for
small fields
\begin{equation}
\rho \rightarrow \frac{\mu m}{\pi } + \frac{|eB|}{2 \pi},
\end{equation}
also one unit of fermion charge. That is, in the
nonrelativistic electron gas the fluxon may be thought of as a
fermion in that it has both the spin and charge of a fermion.
However, the close connection between spin of a fluxon and induced
Chern-Simons term for arbitrary fields that we found in the relativistic
case is lost. This can be traced back to the fact that in the
nonrelativistic case the electron spin is an independent degree of
freedom. In the next section we point out that the spin of the fluxon
does not derive from the ordinary Chern-Simons term, but from a
so-called mixed Chern-Simons term. Such a term is absent in the
relativistic case.

To see how the spin contribution (\ref{textb}) to the magnetic
susceptibility compares to the orbital contribution we evaluate the
$k_0$-integral in the effective action (\ref{nreff}) with $b=0$ to
obtain
\begin{equation}
{\cal L}_{\rm eff} =  \frac{|eB|}{2 \pi} \sum_{n=0}^{\infty}
\sum_{\varsigma =\pm} (\mu - E_{n,\varsigma}) \theta(\mu -
E_{n,\varsigma}).
\end{equation}
The summation over $n$ is easily carried out with the result for small
fields
\begin{equation}
{\cal L}_{\rm eff} = \frac{1}{4 \pi} \sum_{\varsigma =\pm} \left[
\mu_\varsigma^2 m - \frac{(eB)^2}{4 m} \right]
              = \frac{\mu^2 m}{2 \pi} + \frac{(eB)^2}{8 \pi m}
[(2\sigma)^2 -1],                                    \label{nr}
\end{equation}
where $\sigma={1 \over 2}$ and  $\mu_\pm$ is given by (\ref{chemical}).
The first term in the right-hand side of (\ref{nr}), which is
independent of the magnetic field, is the free particle contribution
\begin{equation}
\frac{\mu^2 m}{2 \pi} = -2 \int \frac{d^2 k}{(2\pi)^2} \left(
\frac{k^2}{2m} - \mu \right) \theta \left(\mu - \frac{k^2}{2m} \right).
\end{equation}
The second term yields the low-field susceptibility
\begin{equation}
\chi = (-1)^{2 \sigma +1}  2 \mu_{\rm B}^2 \, \nu_{2D}(0) \left[(2\sigma)^2 -
1)\right].                                            \label{nrchi}
\end{equation}
Equation (\ref{nrchi}) shows that
the ratio of orbital to spin contribution to $\chi$ is different
from the three-dimensional case. Also, whereas a $3D$
electron gas is paramagnetic $(\chi>0)$ because of the dominance of the
spin contribution, the $2D$ gas is  not magnetic $(\chi=0)$ at small
fields since the orbital and spin contributions to $\chi$ cancel.

\subsection{Mixed Chern-Simons term.}

As we have seen above, in the nonrelativistic case there are no
true Chern-Simons. Now, we will present consideration of this
problem in nontrivial background field.

In this section we investigate the origin of the induced spin density
(\ref{41}) we found in the nonrelativistic electron gas. To this end we
slightly generalize the theory (\ref{30}) and consider the Lagrangian
\begin{equation}
{\cal L} =\Psi ^{\dagger}\left[ i\partial _{0} - e A_0 + \mu
-\frac{1}{2m}(i {\bf \nabla} + e {\bf A})^{2}
\right] \Psi +\frac{e}{m}B^{a}\Psi ^{\dagger }\frac{\sigma ^{a}}{2}\Psi.
                                                             \label{43}
\end{equation}
It differs from (\ref{30}) in that the spin source term is omitted, and
in that the magnetic field in the Zeeman term is allowed to point in any
direction in some internal space labelled by latin indices
$a,b,c=1,2,3$. As a result also the spin will have components in this
space. It is convenient to consider a magnetic field whose
direction in the internal space varies in space-time. We set
\begin{equation}
B^{a}(x)=B n^{a}(x),                                       \label{n}
\end{equation}
with $n^a$ a unit vector in the internal space. The gauge potential
$A_\mu$ appearing in the first term of (\ref{43}) still gives $\epsilon_{i j}
\partial_i A^j = B$. Equation (\ref{n}) allows us to make the decomposition
\begin{equation}
\Psi (x)=S(x)\chi (x) \; ; \; \; S^{\dagger }S=1,
\end{equation}
with $S(x)$ a local SU(2) matrix which satisfies
\begin{equation}
{\bf \sigma} \cdot  {\bf n}(x)=
S(x)\sigma ^{3}S^{\dagger }(x).
                                                        \label{46}
\end{equation}
In terms of these new variables the Lagrangian (\ref{43}) becomes
\begin{equation}
{\cal L} =\chi ^{\dagger }\left[ i\partial _{0} - eA_0 -V_{0}
    + \mu - \frac{1}{2m}
    (i {\bf \nabla} + e {\bf A}
    + {\bf V})^{2}\right]\chi +
     \frac{eB }{2m}\chi ^{\dagger}\sigma ^{3}\chi,           \label{47}
\end{equation}
where the $2\times 2$ matrix $V_{\mu }=-iS^{\dagger}(\partial _{\mu}S)$
represents an element of the SU(2) algebra, which can be written in
terms of (twice) the generators  $\sigma^a$ as
\begin{equation}
V_{\mu }=V_{\mu }^{a} \sigma ^{a}.
\end{equation}
In this way the theory takes formally the form of a gauge theory with
gauge potential $V_{\mu }^{a}$. In terms of the new fields the spin
density operator,
\begin{equation}
j_{0}^{a}=\Psi ^{\dagger}\frac{\sigma ^{a}}{2}\Psi,            \label{49}
\end{equation}
becomes \cite{Ko}
\begin{equation}
j_{0}^{a}=R_{ab}\chi ^{\dagger}\frac{\sigma ^{b}}{2}\chi = - \frac{1}{2}
R_{ab} \frac{\partial {\cal L} }{\partial V_{0}^{b}}.   \label{50}
\end{equation}
In deriving the first equation we employed the identity
\begin{equation}
S^{\dagger}({\bf \theta})\sigma ^{a}S
({\bf \theta})=
   R_{ab}({\bf \theta})\sigma ^{b},              \label{51}
\end{equation}
which relates the SU(2) matrices in the $j={1 \over 2}$
representation, $S({\bf \theta})=
\exp({i \over 2} {\bf \theta} \cdot
{\bf \sigma})$, to those in the adjoint representation $(j=1)$,
$R( {\bf \theta})=\exp(i {\bf \theta}
\cdot {\bf  J^{\rm adj}})$.
The matrix elements of the generators in the latter representation are
$\left(J_{a}^{{\rm adj}}\right)_{bc}=-i\epsilon _{abc}$.

The projection of the spin density $j_0^a$ onto the spin quantization
axis, i.e.\ the direction $n^a$ of the applied magnetic field \cite{Ko},
\begin{equation}
{\bf n} \cdot {\bf j}_{0}=
 -\frac{1}{2} \frac{\partial \mbox{$\cal L$}}
     {\partial V_{0}^{3}},                                        \label{54}
\end{equation}
only involves the spin gauge field $V_\mu^3$. So when calculating the
induced spin density $s = \langle {\bf n}
\cdot {\bf j}_{0} \rangle$ we
may set the fields $V^1_\mu$ and $V^2_\mu$ equal to zero and consider the
simpler theory
\begin{equation}
{\cal L} = \sum_{\varsigma = \pm} \chi_{\varsigma}^{\dagger }
\left[ i\partial _{0}
- eA_0^{\varsigma} + \mu_{\varsigma} -
\frac{1}{2m} (i {\bf \nabla} + e
{\bf A}^{\varsigma})^{2}\right]\chi_{\varsigma},
\end{equation}
where the effective chemical potentials for the spin-$\uparrow$ and
spin-$\downarrow$ electrons are given in (\ref{chemical})
and $eA_\mu^\pm = eA_\mu \pm V_\mu^3$. Both components
$\chi_\uparrow$ and $\chi_\downarrow$ induce a Chern-Simons term, so
that in total we have
\begin{eqnarray}
{\cal L}_{\rm CS} &=& \frac{e^2}{2} \epsilon^{\mu \nu \lambda}
( \theta_+ A^+_\mu \partial_\nu
A^+_\lambda + \theta_- A^-_\mu \partial_\nu  A^-_\lambda) \label{total}  \\
           &=&  \frac{\theta_+ + \theta_-}{2} \epsilon^{\mu \nu \lambda} (e^2
A_\mu \partial_\nu  A_\lambda + V^3_\mu \partial_\nu  V^3_\lambda)
+ e (\theta_+ - \theta_-)
\epsilon^{\mu \nu \lambda} V^3_\mu \partial_\nu  A_\lambda,  \nonumber
\end{eqnarray}
where the last term involving two different vector potentials is a mixed
Chern-Simons term. The coefficients are given by
\begin{equation}
\theta_\pm = \frac{1}{2 \pi} {\rm sgn}(eB) N_\pm ,
\end{equation}
assuming that $|eB|>{1 \over 2} |\epsilon_{i j} \partial_i V_j^3|$, so
that the sign of $eB$ is not changed by spin gauge contributions. The
integers $N_\pm$ are the number of filled Landau levels for
spin-$\uparrow$ and spin-$\downarrow$ electrons given by (\ref{npm}).
Since $N_+ -N_- = {\rm sgn} (eB)$, we obtain for the induced spin
density $s$ precisely the result (\ref{41}) we found in the preceeding
section,
\begin{equation}
s = \langle {\bf n} \cdot {\bf j}_{0}
\rangle = -\frac{1}{2} \left.
\frac{\partial {\cal L}_{\rm eff}}
{\partial V_{0}^{3}} \right|_{V_\mu^3 = 0} =
\frac{eB}{4 \pi}.
\end{equation}
The present derivation clearly shows that the induced spin in
the nonrelativistic electron gas originates not from the standard
Chern-Simons term, but from the mixed Chern-Simons term
involving the electromagnetic and spin gauge potential.

The first term in (\ref{total}) is a standard Chern-Simons term, the
combination $\theta_+ + \theta_-$ precisely reproduces the result
(\ref{nrth}) and is related to the induced fermion number density
(\ref{39}).

\section{Trace identity.}
As it was shown in \cite{ni,ni2,ni1} the trace identities
connect Chern-Simons
term and chiral anomaly. These identities may be derived for
Hamiltonians of the form
\be
\label{n61}
H=\left[
\begin{array}{ll}
 m & D \\
 D^{+}& -m
\end{array}
\right]  .
\ee
Here $m$ is a constant, $D$ a differential operator of the form
$D=iP_{i}\partial_{i}+Q(x)$, and $D^{+}$ the hermitian conjugate of $D$.
The $P_{i}$ are constant matrices that satisfy
$P^{+}_{i}P_{j}+P^{+}_{j}P_{i}=2\delta_{ij}$ and
$P_{i}P^{+}_{j}+P_{j}P^{+}_{i}=2\delta_{ij}$ and $Q(x)$ includes all
background fields. It is assumed that these background fields are static,
so
\be
H=H_{0}+m\Gamma^{c}=i\Gamma_{i}\partial_{i}+K(x)+m\Gamma^{c},
\ee
where
\be
\Gamma_{i}=\left[
\begin{array}{ll}
 0 & P_{i} \\
 P^{+}_{i}& 0
\end{array}
\right] ,
\Gamma_{c}=\left[
\begin{array}{ll}
 1 & 0 \\
 0 & -1
\end{array}
\right] ,
K(x)=\left[
\begin{array}{ll}
 0 & Q(x) \\
 Q^{+}(x) & 0
\end{array}
\right] ,
\ee
here $\Gamma$ matrices satisfy the euclidian Dirac algebra
and the operator $H_{0}$ anticommutes with $\Gamma^{c}$.
As a consequence $H^{2}=H^{2}_{0}+m^{2}\ge m^{2}$ and all eigenvalues of
$H$
\be
\label{n610}
H\Psi=H\left[
\begin{array}{l} u\\ v\end{array}\right]
=\lambda\left[
\begin{array}{l} u\\ v\end{array}\right]=\lambda\Psi
\ee
satisfy $\lambda^{2}\ge m^{2}$. Using (\ref{n61}) we obtain the
first-order equations
\be
D^{+}u=(\lambda + m)v,\;\;\; Dv=(\lambda -m)u
\ee
and by iterating, we find
\be
\label{n612a}
DD^{+}u=(\lambda^{2} - m^{2})u,
\ee
\be
\label{n612b}
D^{+}Dv=(\lambda^{2} - m^{2})v.
\ee
If $u$ is a solution of (\ref{n612a}) with eigenvalue $\lambda^{2}-m^{2}=
\chi\ne 0$ then $D^{+}u$ is a solution of (\ref{n612b}) with the
same eigenvalue $\chi$. However, if $u$ is a zero mode of $D^{+}$,
in general it is not a zero mode of $D$. Every solution
of (\ref{n612a}) or (\ref{n612b}) yields two solutions of
(\ref{n610}) if $\lambda\ne \pm m$ and one if $\lambda = \pm m$,
and consequently the Dirac problem (\ref{n610}) is
equivalent to (\ref{n612a},\ref{n612b}).

The fermion number operator has the form  (for discussion on
fermion number definition see beginning of the section 8)
\be
N=\frac{1}{2}\int dx\left[ \Psi^{+}(x),\Psi(x)\right].
\ee
At the time $t=0$ the second quantized fermion field operator can be
expanded as
\be
\Psi(x)=\sum_{n}b_{n}\omega_{n}(x) +\sum_{n}d^{+}_{n}\phi_{n}(x)+
\int dk\left(b_{k}\omega_{k}(x) + b^{+}_{k}\phi_{k}(x)\right),
\ee
where $\omega_{n}(x)$ and $\phi_{n}(x)$ are the positive and negative
energy bound state solutions of the eigenvalue equation
\be
H\psi_{n}=\lambda_{n}\psi_{n},
\ee
and $\omega_{k}(x)$ and $\phi_{k}(x)$ are the positive and negative
energy continuum solutions.
Thus, the fermion number operator can be rewritten as follows
\be
N=N_{0}-\frac{1}{2}\eta_{H},
\ee
where we have defined
\be
N_{0}=\sum_{n}\left[b^{+}_{n}b_{n}-d^{+}_{n}d_{n}\right] +
\int dk\left[b^{+}_{k}b_{k}-d^{+}_{k}d_{k}\right],\nonumber\\
\eta_{H}=\sum_{k} \sgn (\lambda_{k}).
\ee
The summation extends over both the descrete and continuum portions
of the spectrum, and if a continuum spectrum is present,
we rather write as
\be
\label{n128}
\eta_{H}=\int d\lambda\rho_{H}(\lambda) \sgn (\lambda ).
\ee
Here $\rho_{H}(\lambda)$ is the spectral density function of the
hamiltonian $H$, and we may express it in terms of its even and
odd parts:
\be
\rho_{H}(\lambda)=\frac{1}{2}\left[\rho_{H}(\lambda)
+\rho_{H}(-\lambda)\right] + \frac{1}{2}\left[\rho_{H}(\lambda)
-\rho_{H}(-\lambda)\right] =\tau_{H}(\lambda) +\sigma_{H}(\lambda).
\ee
If we substitute it in (\ref{n128}) we obtain
\be
\eta_{H}=\int d\lambda\sigma_{H}(\lambda) \sgn (\lambda )
\ee
since only the odd part of $\rho_{H}(\lambda)$ can contribute
to $\eta_{H}$.
So, $\eta_{H}$ yields the difference in the number of positive
and negative energy eigenstates of the hamiltonian $H$, and
thus it is a measure of its spectral asymmetry. However,
the sum is not absolutely convergent and it needs to be regulated:
the Atiyah--Patodi--Singer $\eta$--invariant of the hamiltonian $H$
is defined by
\be
\eta_{H}(s)=\sum_{k} \sgn (\lambda_{k} ) \vert\lambda\vert^{-s}=
\int d\lambda\sigma_{H}(\lambda) \sgn (\lambda )\vert\lambda\vert^{-s}.
\ee
For a large class of hamiltonians the residue at $s=0$ vanishes, and we
assume that $s=0$ is a regular point of $\eta_{H}(s)$, so we can define
\be
\label{166}
\eta_{H}=\lim_{s\rightarrow 0}\eta_{H}(s)=\sum_{k}\sgn(\lambda_{k})
\equiv \int d\lambda\sigma_{H}(\lambda)\sgn (\lambda).
\ee

We shall now show how the spectral density $\rho_{H}(\lambda)$ of
the Hamiltonian (\ref{n61}) can be represented in terms
of the spectral densities $\rho_{DD^{+}}(\chi)$ and
$\rho_{D^{+}D}(\chi)$ of the operators $DD^{+}$ and $D^{+}D$
respectively. For this we first consider the following Stieltjes
transformation of the even part of $\rho_{H}(\lambda)$:
\be
\label{n618}
\int_{-\infty}^{\infty}d\lambda \rho_{H}(\lambda)
\frac{1}{\lambda^{2}+z^{2}}=
2\int_{\vert m\vert}^{\infty}d\lambda \tau_{H}(\lambda)
\frac{1}{\lambda^{2}+z^{2}}.
\ee
Here $z^{2}$ is an arbitrary complex number which does not belong
to the spectrum of $H$. Introducing the coordinate representation
we obtain
\be
\label{n619}
2\int_{\vert m\vert}^{\infty} d\lambda \tau_{H} (\lambda)
\frac{1}{\lambda^{2}+z^{2}} = \int dx \tr\langle x\vert
\frac{1}{H^{2}+z^{2}} \vert x\rangle \nonumber\\
=\int dx\left( \tr\langle x\vert\frac{1}{DD^{+}+m^{2}+z^{2}}
\vert x\rangle+\tr\langle x\vert\frac{1}{D^{+}D+m^{2}+z^{2}}
\vert x\rangle \right) \nonumber\\
=\int d\chi\left( \rho_{DD^{+}}(\chi)+\rho_{D^{+}D}(\chi)
\right) \frac{1}{\chi+m^{2}+z^{2}}\equiv F(m^{2}+z^{2})
\ee
Comparing (\ref{n618}) with (\ref{n619}) we conclude that
\be
\tau_{H}(\lambda)=\vert\lambda\vert\left(
\rho_{DD^{+}}(\lambda^{2}-m^{2}) +\rho_{D^{+}D}(\lambda^{2}-m^{2})\right).
\ee
Similarly, we find a representation for the odd part of $\rho_{H}(\lambda)$
by considering
\be
\label{n621}
2\int_{\vert m\vert}^{\infty} d\lambda \sigma_{H} (\lambda)
\frac{\lambda}{\lambda^{2}+z^{2}} = \int dx \tr\langle x\vert
\frac{H}{H^{2}+z^{2}} \vert x\rangle \nonumber\\
=\int dx\left( \tr\langle x\vert\frac{m}{DD^{+}+m^{2}+z^{2}}
\vert x\rangle-\tr\langle x\vert\frac{m}{D^{+}D+m^{2}+z^{2}}
\vert x\rangle \right) \nonumber\\
=\int d\chi\left( \rho_{DD^{+}}(\chi)-\rho_{D^{+}D}(\chi)
\right) \frac{m}{\chi+m^{2}+z^{2}}\equiv mG(m^{2}+z^{2}).
\ee
We conclude that the odd spectral density, which contribute to
(\ref{166}), is given by
\be
\sigma_{H}(\lambda)=m\sgn(\lambda)\left[
\rho_{DD^{+}}(\lambda^{2}-m^{2})-\rho_{D^{+}D}(\lambda^{2}-m^{2})\right].
\ee

Since the fermion number is essentially a Mellin transformation of the odd
spectral density, we find the following spectral representation for the
fermion number
\be
\label{172}
N=-m\int_{0}^{\infty} d\lambda\left[
\rho_{DD^{+}}(\lambda^{2}-m^{2})-\rho_{D^{+}D}(\lambda^{2}-m^{2})\right]=
\nonumber\\ =-\frac{1}{2}\int_{0}^{\infty} d\chi\left[
\rho_{DD^{+}}(\chi)-\rho_{D^{+}D}(\chi)\right]
\frac{m}{\sqrt{\chi+m^{2}}}=\nonumber\\
=-\frac{m}{\pi}\int_{0}^{\infty} d\omega G(m^{2}+\omega^{2}),
\ee
where $G(m^{2}+\omega^{2})$ is defined by (\ref{n621}).
We shall now show how the axial anomaly is connected with fermion number.
First notice that
\be
\tr\langle x\vert\frac{m}{DD^{+}+m^{2}+\omega^{2}}
\vert y\rangle-\tr\langle x\vert\frac{m}{D^{+}D+m^{2}+\omega^{2}}
\vert y\rangle =\nonumber\\
=\tr\langle x\vert\Gamma^{c}\frac{m}{H^{2}+\omega^{2}}
\vert y\rangle.
\ee
Further, we get
\be
\label{n87}
\tr\langle x \vert \Gamma^{c}\frac{m}{H^{2}+\omega^{2}} \vert y\rangle=
i\frac{m}{\sigma} \tr\langle x \vert \Gamma^{c}
\frac{1}{H_{0}+i\sigma} \vert y\rangle,
\ee
where $\sigma=\sqrt{m^{2}+\omega^{2}}$.
Let's now consider
\be
\label{n89}
\tr\langle x \vert i\Gamma^{i}\partial_{i} \Gamma^{c}
\frac{1}{H_{0}+i\sigma} -  \Gamma^{c}
\frac{1}{H_{0}+i\sigma} i\Gamma^{i}\partial_{i}\vert y\rangle=\nonumber\\
=2i\sigma\tr\langle x \vert \Gamma^{c}\frac{1}{H_{0}+i\sigma}
\vert y\rangle +\tr\left( [K(y)-K(x)]
\langle x \vert \Gamma^{c}\frac{1}{H_{0}+i\sigma}
\vert y\rangle
\right).
\ee
Combining (\ref{n87}) and (\ref{n89}) we then obtain the
following trace identity
\be
\label{n810}
i\frac{m}{\sigma} \tr\langle x \vert \Gamma^{c}
\frac{1}{H_{0}+i\sigma} \vert y\rangle =\frac{m}{2\sigma^{2}}
[\partial_{x}+\partial_{y}]\tr
\langle x \vert i\Gamma^{i} \Gamma^{c}\frac{1}{H_{0}+i\sigma}
\vert y\rangle +\nonumber\\
+\frac{m}{2\sigma^{2}}\tr\left( [K(y)-K(x)]
\langle x \vert \Gamma^{c}\frac{1}{H_{0}+i\sigma}
\vert y\rangle  \right).
\ee
Notice that  (\ref{n810}) has the structure  of the standard axial
anomaly equation for the DIrac operator $H_{0}+i\sigma$. Thus,
when we take $x\rightarrow y$ limit we need to discuss two cases:

First, if  the space dimension $D$ is odd  the second term on the
right-hand side of (\ref{n810}) vanishes, since
there are no axial anomaly in this case, and the only contribution
to fermion number gives first term (boundary term).

Second, if the space dimension $D$ is even the second term on the
right-hand side of (\ref{n810}) gives the axial anomaly and
taking in mind (\ref{172}) we get for fermion number
\be
N=-\frac{1}{2\pi}\int_{0}^{\infty}d\omega\frac{m}{m^{2}+\omega^{2}}
\left( 2T_{D}+ \int dS^{i}\tr\langle x\vert i\Gamma^{i}\Gamma^{c}
\frac{1}{H_{0}+i\sigma}\vert x\rangle\right),
\ee
here $T_{D}$ is the Pontryagin index of the background gauge fields
that arises from the space integral of anomaly term. There is also
boundary term, which vanishes for  a trivial gauge background.

Let us consider 2-dimensional case, for the Hamiltonian
\be
H_{0}=-i\sigma^{2}\frac{d}{dx}+\sigma^{1}\phi (x).
\ee

Since this Hamiltonian
can be interpreted as a one-dimensional Dirac operator, and since there
are no anomalies in one dimension, the fermion number is given by
\be
N=-\frac{1}{2\pi}\int_{0}^{\infty}d\omega\frac{m}{m^{2}+\omega^{2}}
\int_{-\infty}^{+\infty} dx\frac{d}{dx}
\tr\langle x\vert i\Gamma^{1}\Gamma^{c}
\frac{1}{H_{0}+i\sigma}\vert x\rangle=\nonumber\\
=
\frac{1}{2\pi}\int_{0}^{\infty}d\omega\frac{m}{m^{2}+\omega^{2}}
\Bigl[
\tr\langle \infty\vert \sigma^{1}
\frac{1}{H_{0}+i\sigma}\vert \infty\rangle -\nonumber\\
-\tr\langle -\infty\vert \sigma^{1}
\frac{1}{H_{0}+i\sigma}\vert -\infty\rangle \Bigr]
\ee
where we have used the representation $\Gamma^{1}=\sigma^{2}$ and
$\Gamma^{c}=\sigma^{3}$ of the Dirac algebra.
We assume that the soliton field $\phi (x)$ has the asymptotes
$\phi (\pm\infty)=\hat{\phi}_{\pm}$. Taking in mind that
\be
\tr\langle \pm\infty\vert \sigma^{1}
\frac{1}{H_{0}+i\sigma}\vert \pm\infty\rangle =\nonumber\\
=\tr\langle \pm\infty\vert
\frac{1}{-\sigma^{3}\partial-\phi(x) -i\sigma^{2}\sigma^{3}\sqrt{m^{2}
+\omega^{2}}}\vert \pm\infty\rangle =\nonumber\\
2\tr\langle \pm\infty\vert
\frac{\phi(x)}{\partial^{2}-\phi^{2}(x) -m^{2}-\omega^{2} }\vert
\pm\infty\rangle =\nonumber\\
=-2\hat{\phi}_{\pm}\int_{-\infty}^{+\infty}\frac{dk}{2\pi}
\frac{1}{k^{2}+\hat{\phi}^{2}_{\pm} +m^{2}+\omega^{2} },
\ee
we find for fermion number
\be
N=-\frac{1}{2\pi}\left[{\rm arctan}\left[\frac{\hat{\phi}_{+}}{m}\right]
-{\rm arctan}\left[\frac{\hat{\phi}_{-}}{m}\right]
\right].
\ee
This result \cite{ni3,ni1} is obtained without the use of any specific
soliton profile. However, at the presence of nonzero density $(\mu)$,
fermion number  depends not  only on asymptotic properties of soliton
profile, but also  it depends on the local properties of the soliton
profile such as the width of the soliton \cite{soni}.

The result for fermion number at the presence of density and temperature
one can find in \cite{soni}. There were  studied the Hamiltonian \cite{wil}
\be
H_{0}=\sigma^{2}\frac{d}{dx}+\sigma^{1}\phi (x)+\sigma^{3}\epsilon.
\ee
This Hamiltonian has the following  positive- and negative-energy
continuum solutions and a bound state solution (assuming a soliton profile
which has only one bound state)
\be
\psi_{k\alpha}&=& \left[
\begin{array}{l} {[ (\alpha E +\epsilon ) / 2\alpha E ]^{1/2} u_{k}} \\
{[ 2\alpha E (\alpha E +\epsilon ) ]^{1/2} (\partial_{x} +\phi ) u_{k}}
\end{array}
\right],\nonumber\\
\psi_{s}&=&N_{0}\left[\begin{array}{l}{\exp [ -\int^{x}dx' \phi(x')]}\\
{0}\end{array}\right],
\ee
where $N_{0}$ is a normalization factor, $\alpha=\pm 1$ distinguishes
positive and negative energy solutions.
In the ground state the soliton charge is defined as
\be
\label{so7}
Q=\int_{-\infty}^{\infty}dx\sum \left[\rho_{i}^{s}(x)-\rho_{i}^{0}(x)
\right],
\ee
where $\rho_{i}^{s}(x)$ and $\rho_{i}^{0}(x)$ are the fermion number density
at a point $x$ in the presence and absence of the soliton, due to occupied
state $i$.

The generalization of  (\ref{so7})  to finite $\mu$ and $T$
is straightforward since we have a noninteracting sea of fermions
\be
Q(\mu,T)=\int_{-\infty}^{\infty}dx\sum \left[\rho_{i}^{s}(x)-\rho_{i}^{0}(x)
\right] n(\epsilon_{i}-\mu ),
\ee
where $n(\epsilon -\mu)=\left[ \exp[\beta(\epsilon -\mu)] + 1\right]^{-1}$
is the Fermi distribution function.
Thus, substitution of $\rho$ yields
\be
\label{so9}
Q=\int_{-\infty}^{\infty}dx\sum_{\alpha=\pm 1}
\int_{-\infty}^{\infty}\frac{dk}{2\pi} ( |u_{k}^{s}|^{2} -|u_{k}^{0}|^{2})
n(\alpha E -\mu)\nonumber\\ + \sum_{\alpha=\pm 1}
\int_{-\infty}^{\infty}\frac{dk}{2\pi}\left[
\frac{ (\partial_{x}|u_{k}^{s}|^{2} +2|u_{k}|^{2}\phi)
|_{x=-\infty}^{x=+\infty} }{4\alpha E (\alpha E +\epsilon)}
\right] n(\alpha E -\mu) + n(\epsilon -\mu).
\ee
The square bracket in the second term of the above expression
can be simplified further using
$(\partial_{x}|u_{k}^{s}|^{2} +2|u_{k}|^{2}\phi)
|_{x=-\infty}^{x=+\infty}=2$.
For $T=0$ and $\mu=0$ the first term is easily evaluated using the
completeness properties of $u_{k}$. But for finite $\mu$ we have to choose
a soliton profile. So we take
$\phi (x)=\phi_{0} \th (\phi_{0} x)$, for which the eigenfunctions
$u_{k}^{s}(x)$ are known exactly \cite{wil} to be
\be
u_{k}(x)=-\exp (ikx)\left[
\frac{\th \phi_{0}x - (ik/\phi_{0})}{1+ (ik/\phi_{0})}
\right].
\ee
Substitution of the $u_{k}$ in (\ref{so9}) yields
\be
Q(\mu,T)=-2\phi_{0}\sum_{\alpha=\pm 1}
\int_{-\infty}^{\infty}\frac{dk}{2\pi}
\frac{n(\alpha E -\mu)}{k^{2}+\phi_{0}^{2}}+\nonumber\\
+2\phi_{0}\sum_{\alpha=\pm 1}
\int_{-\infty}^{\infty}\frac{dk}{2\pi}
\frac{n(\alpha E -\mu)}{2\alpha E (\alpha E + \epsilon)}+
n(\epsilon -\mu).
\ee
In particular this integrals can be evaluated exactly for zero temperature
and finite $\mu$ to get \cite{soni}
\be
Q(\mu , 0)=-\sgn(\mu)Q_{0}(\epsilon) -\theta(\mu)G(k_{F},\epsilon)+
\theta(-\mu)G(k_{F}, -\epsilon)
\ee
for $|\mu | > m$, where
\be
Q_{0}(\epsilon)= -\frac{1}{\pi} {\rm tan}^{-1}
\left[\frac{\phi_{0}}{\epsilon}\right],\nonumber\\
G(k_{F}, \epsilon)= \frac{w}{\pi} {\rm tan}^{-1}
\left[\frac{\phi_{0} {\rm tan}[\frac{1}{2}{\rm tan}^{-1}(k_{F}/m)]}
{m+\epsilon}\right],\nonumber\\
k_{F}=(\mu^{2}-m^{2})^{1/2}\;\;, \;\;\;\;
m=(\phi_{0}^{2} +\epsilon^{2})^{1/2}.\nonumber
\ee

As we have seen above, the boundary term is dependent on
soliton profile at finite density.
Thus, the generalization of the trace
identities on finite density and temperature is hardly possible
because of nontopological part of it.

At finite temperature and zero density
trace identity still holds  and one has for fermion number \cite{ni,ni1}
\be
\label{in}
\langle N\rangle_{\beta}=-\frac{1}{2\beta}\sum_{-\infty}^{+\infty}
\frac{m}{m^2 +\omega^{2}_{n}}\nonumber\\
\left(\int dx ({\rm anomaly}) +
\int dx \partial_{i}\tr\langle x |i\Gamma_{i}\Gamma^{c}
\frac{1}{H_{0}+i\sqrt{m^2 + \omega^{2}_{n}}}\rangle\right).
\ee

Now we'll prove that chiral anomaly doesn't depend on temperature
at any even dimension.
The second term at left hand side of (\ref{in}) is a surface term,
which doesn't
contribute to topological part of the trace identity \cite{ni,ni1}.
So , we won't consider nontopological part of the trace identity,
i.e. nontopological part of fermion density and  surface term.
Thus for topological part trace identity takes the form
\be
\label{in1}
\langle N\rangle_{\beta}^{\rm topological}
=-\frac{1}{2\beta}\sum_{-\infty}^{+\infty}
\frac{m}{m^2 +\omega^{2}_{n}}\left(\int dx ({\rm anomaly})
\right).
\ee
The result for left hand side of  eq.(\ref{in1}) we know
in arbitrary odd dimension.  Really, from (\ref{kon}) we have
\be
\label{dl}
\langle N\rangle_{\beta}^{\rm CS}=
\langle N\rangle_{\beta}^{\rm topological}
=\frac{\delta I^{{\rm C.S}}_{eff}}{g\delta A_{0}}.
\ee
By using the fact that
\be
\frac{1}{2\beta} \sum_{n=-\infty}^{+\infty}
\frac{m}{\omega_{n}^2+m^2} =\frac{1}{4}\frac{\sh (\beta m)}
{1+\ch (\beta m)},
\ee
one can see that the only possibility to reconcile left and right
sides of eq.(\ref{in1}) is to put temperature independence of
anomaly. Thus, we proof that axial anomaly doesn't depend on
temperature in any even-dimensional theory.

Moreover, now we can generalize trace identity on arbitrary
finite density.
Really, taking in mind (\ref{kon}) and (\ref{dl}) one can see
\be
\label{odd}
\langle N\rangle_{\beta , \mu}^{\rm CS}= -
\frac{1}{4}\th (\beta m)\frac{1}{1+\ch (\beta\mu)/\ch (\beta m)}
\int dx \left( {\rm anomaly} \right),
\ee
where $\langle N\rangle_{\beta , \mu}^{\rm CS}$ --
odd part of fermion number in $D$-dimensional theory
at finite density and temperature,  $\left( {\rm anomaly} \right)$ --
axial anomaly in $(D-1)$-dimensional theory.
On the other hand, as we have seen above, the anomaly doesn't
depend on $\mu$  in 2 and 4 dimensions and doesn't
depend on $T$ in any even-dimensional theory.
Our comprehension of the problem allows us to
generalize these on arbitrary even dimension.
Indeed, anomaly is  the result of  ultraviolet regularization,
while $\mu$ and $T$ don't effect on ultraviolet
behavior of a theory. Taking in mind (\ref{odd})  and
that at finite density
\be
\frac{1}{2\beta} \sum_{n=-\infty}^{+\infty}
\frac{m}{\omega_{n}^2+m^2} =\frac{1}{4}
\th (\beta m)\frac{1}{1+\ch (\beta\mu)/\ch (\beta m)}
\ee
we can identify
$\langle N\rangle_{\beta , \mu}^{\rm topological}$ and
$\langle N\rangle_{\beta , \mu}^{\rm CS}$.
So, we get generalized on finite density trace identity
\be
\label{g}
\langle N\rangle_{\beta ,\mu}^{\rm CS}=
\langle N\rangle_{\beta , \mu}^{\rm topological}=
-\frac{1}{2\beta}\sum_{-\infty}^{+\infty}
\frac{m}{m^2 +\omega^{2}_{n}}\left(\int dx ({\rm anomaly})\right).
\ee

Let us take, for example, 3--dimensions. We know that chiral anomaly
in 2-dimensions has the form
\be
\int d^{2}x\frac{e}{4\pi^{2}}\epsilon^{ij}F_{ij},
\ee
substituting this in (\ref{g}) we'll get for fermion number
\be
\langle N\rangle_{\beta ,\mu}^{\rm CS}=\frac{e}{16\pi^{2}}
\th (\beta m)\frac{1}{1+\ch (\beta\mu)/\ch (\beta m)}
\int d^{2}x \epsilon^{ij}F_{ij}.
\ee
Covariantizing fermion number we get for the Chern-Simons term in action
\be
I^{\rm CS}_{eff} =\th (\beta m)\frac{1}{1+\ch (\beta\mu)/\ch (\beta m)}
\frac{g^2}{16\pi}
\int_{x} e^{\mu\nu\alpha} \tr\left(
A_{\mu}F_{\nu\alpha}\right).
\ee

Really, we've seen that only zero modes contribute to $P$-odd part
in contrast to $P$-even part which is contributed by all modes.
Therefore, index theorem and trace identities hold only
for parity odd part of fermion number.
Thus, the main result of this section is eq.(\ref{g})
which connects Chern-Simons term and chiral anomaly in arbitrary dimensional
theory at finite density and temperature.

\section{Conclusions.}
Thus, it is obtained finite temperature and density influence
on Chern-Simons term generation in any odd dimensional theory as for
abelian, so as for nonabelian case.
It is very interesting that  $\mu^2 = m^2$ is the
crucial point for Chern-Simons at zero temperature.
Indeed, it is clearly seen from (\ref{kon}) that when $\mu^2 < m^2$
$\mu$--influence disappears and we get the usual Chern-Simons term
$I^{{\rm CS}}_{eff}= \pi W[A].$
On the other hand when $\mu^2 > m^2$
the Chern-Simons term disappears because of non--zero
density of background fermions.
The coefficient at Chern-Simons term is the same in any odd dimension.
It must be stressed that
at $m=0$  even negligible density or temperature,
which always take place in any
physical processes, leads to vanishing of the parity anomaly.

It is shown that the chiral anomaly doesn't influenced by medium
effects such as chemical potential and temperature in any
even dimensional theory. Moreover, even if we introduce
conservation of chiral charge on quantum level, the chiral
anomaly arises and isn't affected.

The appearance of Chern-Simons number in even-dimensional theories is
discussed under two type of constraints. So, it is
shown both for conserved charge, i.e. finite density of the
background fermions, and for conserved axial charge what
corresponds to conservation of the left(right)-handed
fermions asymmetry in the background.

The topological part of the trace identity is generalized
on finite density. Thus, the connection between Chern-Simons term
and chiral anomaly at finite density and temperature
is obtained in arbitrary dimensional theory.

In conclusion we would like to touch the problem, which
have attracted recently a wide interest \cite{lee,des}.
This is  gauge invariance of the effective action under
large gauge transformations.  Really, the Chern-Simons term
coefficient has to be {\it "topologically quantized"}
for gauge invariance of the effective action
under large  gauge transformations. But as we have seen above
(\ref{z}),(\ref{z1}),(\ref{m3}),
even in nonperturbative calculations of the Chern-Simons in
even dimensions (due to existence of the chiral anomaly),
it gets chemical potential
(temperature) as a coefficient, which is not an integer function.
This fact is hardly understandable. One can treat these that
density or temperature just break invariance under large gauge
transformations, leaving action invariant under local ones.
On the other hand,  one can hope that the whole effective
action will be gauge invariant \cite{lee,des}. But, for example,
essentially nonperturbative and simple calculations in one
dimension \cite{lee} gives not understandable contradiction
between fermion number and effective action. That is the fermion number
here is not a functional derivative of the effective action, what
is very strange.

The amazing fact is that at zero temperature and finite density
the Chern-Simons term coefficient does not break gauge invariance.
Indeed, theta function gives us $0$ or $1$ as the coefficient, and
we have two topological domains $\mu^2 > m^2$ and
$\mu^2 < m^2$ connected by large gauge transformations.

Thus, this area is yet an open field for research.

\end{document}